\documentclass[final, times, twocolumn]{elsarticle}
\usepackage[margin=1.5cm]{geometry}
\let\OLDthebibliography\thebibliography
\renewcommand\thebibliography[1]{
  \OLDthebibliography{#1}
  \setlength{\parskip}{0pt}
  \setlength{\itemsep}{0pt plus 0.3ex}
}
\usepackage[subrefformat=parens,labelformat=parens]{subfig}
\usepackage{epsfig}
\usepackage{soul,color}
\usepackage{amssymb}
\usepackage{amsmath}
\usepackage{textcomp}
\usepackage{graphicx}
\usepackage{subfig}
\PassOptionsToPackage{hyphens}{url}\usepackage[hidelinks]{hyperref}
\usepackage{multirow}
\usepackage{array}
\usepackage{siunitx} 
\usepackage{booktabs}
\usepackage{makecell}
\usepackage{enumitem}
\usepackage{soul}

\biboptions{numbers,sort&compress}

\definecolor{green}{rgb}{0.0,0.6,0.0}
\definecolor{hanpurple}{rgb}{0.32,0.09,0.98}
\definecolor{blue}{rgb}{0,0,1}
\definecolor{red}{rgb}{1, 0, 0}

\newcommand{\tck}{\textcolor{black}}
\makeatletter
\renewcommand\subsection{\@startsection{subsection}{2}{\z@}%
  {12pt plus 1pt minus .2pt}%
  {6pt plus .2pt}%
  {\normalfont\normalsize\bfseries}} 
  
\renewcommand\subsubsection{\@startsection{subsubsection}{3}{\z@}%
  {12pt plus 1pt minus .2pt}%
  {6pt plus .2pt}%
  {\normalfont\normalsize\bfseries}}  
\makeatother

\begin{document}

\begin{frontmatter}

\author[uvm,gund]{Jayashree Yadav\corref{cor1}}
\author[nwab]{Ingemar Mathiasson}
\author[rubenstein]{Bindu Panikkar}
\author[uvm,gund]{Mads Almassalkhi}

\cortext[cor1]{Corresponding author: jyadav@uvm.edu}

\address[uvm]{Electrical and Biomedical Engineering, University of Vermont, Burlington, VT 05401, USA}
\address[gund]{Gund Institute for Environment, University of Vermont, Burlington, VT 05401, USA}
\address[nwab]{Northwest Arctic Borough, Kotzebue, AK 99752, USA}
\address[Rubenstein]{Rubenstein School of the Environment and Natural Resources, University of Vermont, Burlington, VT 05401, USA}

\title{{Community-Centric Multi-Criteria Assessment Framework for Energy Transition}}

\begin{abstract}
The transition to low-carbon energy systems demands comprehensive technical, economic, environmental, and social evaluation tools. While numerous studies address specific aspects of energy transition, few provide an integrated framework to capture the full spectrum of impacts.
This work developed a community-collaborative assessment framework that integrates intelligent energy devices with optimization-based coordination of energy assets. The proposed framework uses techno-economic, environmental, and social criteria to evaluate transition pathways.
A detailed case study is performed for a remote community in Alaska to assess its applicability, where the feasibility of renewable energy transitions remains underexplored. Three distinct pathways, including heat pump and battery integration, resource coordination, and expanded community solar PV, are analyzed using a year-long dataset of demand, renewable energy, and transformer data. The analysis revealed that using heat pumps lowers the overall energy costs by 30\% and carbon emissions by 28\%. In addition, the share of the population spending more than 10\% of their income on energy falls from 74\% in the existing scenario to 40\% with heat pump adoption, indicating significant affordability improvements. 
By combining a general, community-centric assessment framework with a data-driven case study, this work offers a practical tool for utilities, community stakeholders, and policymakers to work toward equitable and sustainable energy transitions.
\end{abstract}




\end{frontmatter}



\section{Introduction}
\label{sec1}

\begin{table}[h]
\textbf{Notation}\\
   \begin{tabular}{lp{6.5cm}}
    $k$ & Discrete time instant\\
    $h$ & House index \\
    $\tau$ & Transformer index \\
    $\text{T}$ & Set of transformers, indexed by $\tau$. \\
    $\mathcal{H}_\tau$ & Set of houses connected to transformer $\tau$, indexed by $h$. \\
    $P_\tau[k]$ & Transformer capacity for $\tau \in \text{T}$ \\
    $D_{\tau}[k]$ & Base power demand of transformer $\tau$ at time step $k$. \\
    $P^{\text{HP}}_{\tau, h}[k]$ & Power consumption of the heat pump for house $h$ connected to transformer $\tau$ at time step $k$. \\
    $P^\text{g}$& Power supplied by the generator \\
    $P^{\text{PV}}$ & Power generated by solar PV \\
    $P^\text{b}_\text{c}$ & Battery charging power \\
    $P^\text{b}_\text{d}$ & Battery discharging power \\
    $T^\text{a}_{\tau, h}$& Indoor temperature of house $h$ connected to transformer $\tau$. \\
    $T^\text{m}_{\tau, h}$& Thermal mass temperature of house $h$ connected to transformer $\tau$ \\
    $T^{\text{o}}$ & Outdoor temperature. \\
    $T^{\text{o}}_\text{min}$ & Heat pump temperature cutoff. \\
    $T^\text{a}_{\text{min}}, T^\text{a}_{\text{max}}$ & Minimum and maximum allowable indoor temperatures, respectively. \\
    $P^\text{HP}_{\text{min}}, P^\text{HP}_{\text{max}}$ & Minimum and maximum heat pump power, respectively. \\
    $\Delta t$ & Time step duration. \\
    $C^\text{a}, C^\text{m}$& Thermal capacities for air and thermal mass, respectively. \\
    $U^\text{a}$& Thermal transmittance coefficient for air. \\
    $H^\text{m}$& Thermal coupling coefficient between air and thermal mass. \\
    $Q^\text{oil}$& Thermal energy generated by heating oil.\\
 $\eta^\text{b}$&Battery efficiency\\
 \end{tabular} 
\end{table}

\begin{table}[h]
\textbf{Acronyms}\\
   \begin{tabular}{lp{6.5cm}}
    AEA & Alaska Energy Authority\\
    AVEC & Alaska Village Electric Coop\\
    ACEP & Alaska Center for Energy and Power\\
    CEA & Customer Energy Assistance\\
    COP& Coefficient of performance of the heat pump \\
    DOE & Dept. of Energy\\
    EIA & Energy Information Authority\\
    ETP & Equivalent Thermal Parameter\\
    IUF & Infrastructure Upgrade Factor\\
    NAB & Northwest Arctic Borough\\
    HP & Heat pump\\
    TEES & Techno-Economic, Environmental, and Social\\
    TP & Transition Pathway \\
    PV & Photovoltaic \\
    PCE & Power Cost Equalization \\
    MCDA & Multi-Criteria Decision Analysis\\
    NREL & National Renewable Energy Laboratory\\
   \end{tabular}
\end{table}

{The transition of energy systems from traditional fossil fuels to non-conventional resources such as wind, solar, and biomass is occurring globally \citep{gielen2019role}. The transition pathways may include integrating renewable energy technologies, deploying energy storage systems, improving efficiency, or implementing demand-side management and control strategies.} To determine the most effective transition pathway for a given energy system, various assessment frameworks have been developed and evaluated using different indicators.
\citet{mckenna2018future} examined the feasibility of hydrogen production at regional and local energy systems using indicators such as investment cost, electricity prices, and plant utilization. 
\citet{ibagon2023techno} developed a techno-economic tool for regional hydrogen hubs in Argentina, integrating operational costs, incentives, and energy efficiency. Similarly, \citet{bhandari2021hydrogen} and \citet{osei2024techno} focused on decentralized hydrogen production in national grid-connected systems using the levelized cost of hydrogen (LCOH). Other works explored rooftop solar and electric vehicle integration in urban systems \citep{chang2022energy}, or biomass energy parks with less defined spatial context \citep{van2013techno}. These studies have focused only on techno-economic assessments, which are insufficient to evaluate the real energy transitions.
 
Recent studies have therefore adopted more holistic frameworks that consider social, political, and environmental dimensions. For instance, \citet{cherp2018integrating} proposed a meta-theoretical model combining techno-economic, socio-technical, and political perspectives. \citet{bolwig2020climate} assessed the role of social acceptance in determining the viability of onshore wind projects. \citet{heras2020social} incorporated indicators such as human wellbeing, unemployment, and population density into site selection. \citet{garcia2019measuring} and \citet{tladi2024assessing} evaluated transitions through GDP, employment, and environmental metrics. \citet{vanegas2022beyond} applied a multi-criteria framework considering techno-economic, environmental, and social indicators for offshore wind development. Many reviews highlighted institutional and governance factors to shape energy transition outcomes \citep{zhang2022green, markard2018next, li2025multidimensional}. In Alaska, \citet{holdmann2022critical} and \citet{allen2016sustainable} explored off-grid community renewable energy pathways, focusing on planning, governance, and reliability. \citet{heleno2024resilient} used mixed-integer linear programming to optimize microgrid configurations for rural locations. 

Despite their breadth, most of these studies are still within top-down planning paradigms, focused on grid-connected national or regional systems, often with limited attention to community-level dynamics. \citet{trueworthy2024will} applied community-driven design frameworks to wave energy development. To highlight the importance of bottom-up, context-sensitive strategies for advancing equitable and effective transitions in underserved areas, where standard techno-economic metrics may fall short.

Few studies have focused on community-driven multi-criteria analysis in the remote Arctic region. These communities face unique technical, economic, and social barriers to renewable energy adoption \citep{arriaga2013renewable} and often rely on isolated diesel microgrids, resulting in high electricity costs, supply vulnerability, and elevated greenhouse gas emissions \citep{holdmann2022critical}. Heating demand in these regions can exceed electricity demand by 1.5 to 2 \citep{nwab2022}, increasing the dependence on imported fuels. Additionally, extreme daylight variation, supply chain constraints, and subsidy structure complicate the system planning. These conditions necessitate a community-centered assessment that integrates technical, economic, environmental, and social factors, framed from the bottom up, to reflect community needs and aspirations. 

In this work, we developed a structured assessment framework to evaluate energy transition pathways using Techno-Economic, Environmental, and Social (TEES) metrics using Multi-Criteria Decision Analysis (MCDA). We further applied this framework to an actual remote Alaskan community as a case study. {MCDA has emerged as an important tool for navigating such multi-dimensional energy planning contexts \citep{wang2009review}. Popular methods include the Analytic Hierarchy Process (AHP), Technique for Order of Preference by Similarity to Ideal Solution (TOPSIS), Preference Ranking Organization Method for Enrichment Evaluation (PROMETHEE), and the Compromise Ranking Method (VIKOR), with fuzzy logic often incorporated to handle uncertainty in weights and criteria \citep{kumar2017review}. AHP is the most widely adopted method, and equal-weighting is frequently used for assigning indicator importance \citep{wang2009review}.}



\tck{The key contributions of this manuscript are as follows:}
\vspace{-0.25cm}
\begin{itemize}
    \setlength{\itemsep}{0pt}
    \setlength{\parskip}{-0.5pt}
    \item Develops a comprehensive multi-criteria assessment framework to evaluate energy transition pathways that can integrate intelligent energy devices and optimization-based coordination of energy assets. 
    \item Applies the framework to a detailed case study of a remote community in Alaska, evaluating the techno-socio-economic impacts of three distinct energy transition pathways, including heat pump and battery integration, resource coordination, and expanded community solar PV, compared to the community’s baseline (today’s) system.
    \item Demonstrates that a dual heating system, combining conventional heating with electric heat pumps, is more cost-effective in extreme cold climates than relying solely on heat pumps.    
\end{itemize}  

This article is organized as follows: Section~\ref{sec2.0} introduces the transition pathway assessment framework in detail. The framework employs an optimization technique to capture changes in demand and the dispatch of energy resources, and evaluates the techno-economic, environmental, and social impacts of each transition pathway. 
Section~\ref{sec2} presents the framework's application to a remote community in Alaska. Section~\ref{sec3} provides time series results of a system for different transition pathways. In section~\ref{sec4}, technological, economic, environmental, and social indices for each transition pathway are discussed, along with the framework's applicability across different systems. Section 6 concludes the article.

\section{Transition Pathways Assessment Framework}
\label{sec2.0}
Transition pathways refer to strategies or interventions aimed at shifting from conventional energy systems to more sustainable configurations. Each pathway represents a distinct combination of technologies, operational practices, and policy mechanisms intended to reduce reliance on fossil fuels while enhancing system sustainability and resilience.
\begin{figure}[t!]
    \centering
    \includegraphics[width=0.9\linewidth]{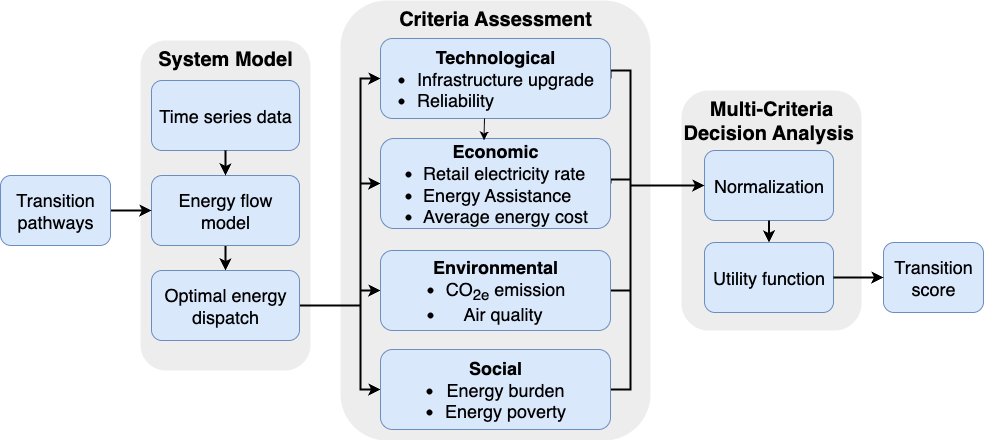}
    \caption{\tck{Multi-criteria assessment framework for evaluating energy transition pathways}}
    \label{analysis_methods}
\end{figure}

Figure~\ref{analysis_methods} presents a multi-criteria assessment framework designed to evaluate energy transition pathways. The framework operates in three stages: (i) system modeling, (ii) criteria assessment, and (iii) MCDA.
The system modeling stage involves simulating energy flows based on time-series data such as energy demand, renewable energy generation, etc. For each transition scenario, this data is fed to an energy flow model to determine the optimal dispatch of available energy resources, such as generators, storage systems, or other technologies. The resulting dispatch outputs are passed on to the second stage. The criteria assessment block evaluates each pathway across four key dimensions, technological, economic, environmental, and social, using a set of indicators shown in Figure~\ref{analysis_methods}. In the final stage, an MCDA approach is employed to aggregate the criteria into a composite transition score. Each indicator is first normalized, and then combined using a utility function to reflect the relative importance of each criterion.

\subsection{System Model}
\label{sys_model_generic}
\begin{figure}[t!]
    \centering
   {\includegraphics[trim = {1cm, 0.5cm, 1.5cm, 0cm},clip, width=0.9\linewidth]{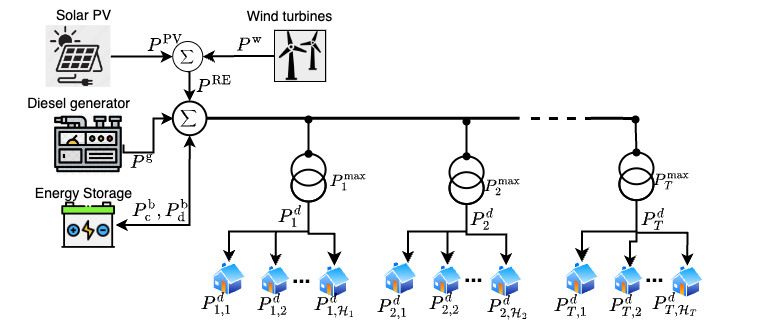} \label{Gen_BD_a}}
    \caption{\tck{Electrical system block diagram of a remote energy community}}
    \label{Gen_BD}
\end{figure}

In this section, we present the model of a variable distribution system in a community, including generation sources and supporting components (e.g., transformers, batteries), to enable the distribution of electricity to end-use loads. 
The generation sources may include diesel generators, solar PV, wind turbines, or other technologies serving residential and/or commercial consumers. Figure~\ref{Gen_BD} illustrates the block diagram of such an electrical distribution system. 

\subsubsection{Energy Flow Model}
The electric distribution system is modeled as an energy flow model. The model assumes a lossless distribution system, with power balance equations governing the energy flow, as expressed in \eqref{EF}. 
Equation~\eqref{EF1} represents the power balance at each time step $k$, where the total demand is met by power generation from diesel ($P^\text{g}$), renewable energy ($P^\text{RE} = P^\text{PV} + P^\text{w} $), and battery discharge ($P^\text{b}_\text{d}$), while battery charging ($P^\text{b}_\text{c}$) is treated as a load. 
Equation~\eqref{EF2} indicates the demand at transformer $\tau$ as the sum of individual household loads $P^\text{d}_{\tau, h}$ connected to it, whereas \eqref{EF3}-\eqref{EF5} are operational constraints. The sets $K$, $\text{T}$, and $\mathcal{H}_\tau$ represent discrete time instants, transformers in the system, and houses connected to transformer $\tau$, respectively.
The power flow components, $P^\text{g}$, $P^\text{RE}$, $P^\text{b}_\text{d}$, $P^\text{b}_\text{c}$, and $P^\text{d}$, are determined through individual component models, optimal dispatch algorithms, and coordination strategies. 
\begin{subequations}
    \begin{align}
        \sum_{\tau=1}^{T} P^\text{d}_\tau[k] &=P^\text{g}[k] + P^\text{RE}[k] + P^\text{b}_\text{d}[k] - P^\text{b}_\text{c}[k]  \quad \forall k \in K   \label{EF1}\\
        P^\text{d}_\tau[k] &= \sum_{h=1}^{\mathcal{H}_\tau} P^\text{d}_{\tau, h}[k] \quad \forall \tau \in \text{T}, \quad \forall k \in K\label{EF2}\\  
        0 &\leq P^\text{g}[k] \leq P^\text{g}_\text{max}[k] \quad \forall k \in K \label{EF3}\\
        0 &\leq  P^\text{b}_\text{c}[k] \leq  P^\text{b}_\text{max}\quad \forall k\in K\label{EF4}\\
        0 &\leq  P^\text{b}_\text{d}[k] \leq P^\text{b}_\text{max}  \quad \forall k\in K\label{EF5}
    \end{align}
    \label{EF}
\end{subequations}

\begin{figure}[htbp]
    \centering
    \includegraphics[width=0.3\textwidth]{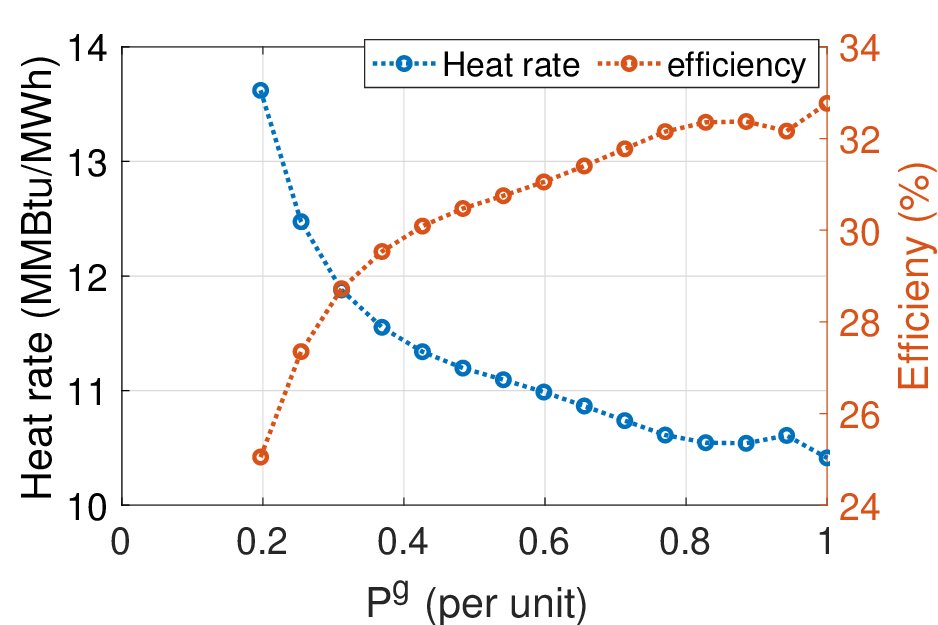}
    \caption{Generator heat rate and efficiency cost curve provided in ~\cite{nrel_data}.}
    \label{gen_costcurve_generic}
\end{figure}

\noindent \textit{a) Diesel Generator Model ($P^\text{g}$)}\\
The diesel generator is modeled using a cost curve, which defines the relationship between fuel consumption cost and power generation. This curve is derived from the generator heat rate curve, indicating how efficiently the generator converts fuel energy into electrical energy. Figure~\ref{gen_costcurve_generic} illustrates the generator heat rate and efficiency curves as per unit power output functions, as provided in an NREL data catalog~\citep{nrel_data}. The generator cost curve is then computed from the heat rate curve using the methodology outlined in Chapter 11, Section 7 of~\cite{bergen_power_2000}.

Based on the specific heat rate data illustrated in Figure~\ref{gen_costcurve_generic}, the total heat input (heat rate multiplied by power output) exhibits a highly linear relationship with respect to the per-unit power output ($P^g$). Consequently, when scaled by a constant fuel price, the fuel consumption cost curve ($C_{P^g}$) for this generator is accurately approximated as a linear function \eqref{Pg_cost}.
\begin{align}
    C_{P^g} = \alpha P^g + C,
    \label{Pg_cost}
\end{align}
where $\alpha$ represents marginal cost of power generation (\$/kWh) and $C$ is the no-load fuel cost (\$/h).

\vspace{0.3cm}
\noindent\textit{b) Renewable Energy Injection ($P^\text{RE}$)}\\
Renewable energy generation is incorporated as available data or through models based on irradiance or wind flow when data is unavailable. In this work, power generation data for existing renewable energy resources were available and directly utilized. For scenarios with increased renewable capacity, the generated power is scaled as follows, 
\begin{equation}
    P^{RE}_\text{new}[k] = P^{RE}_\text{base}[k]  M \quad \forall k \in K,
    \label{pv_power}
\end{equation}
where $P^{RE}_\text{new}$ and $P^{RE}_\text{base}$ represent the scaled and original power generated from renewable sources, respectively, and $M$ is the capacity scaling factor.

\vspace{0.3cm}
\noindent\textit{c) Battery Model ($P^\text{b}$)}\\
The battery is modeled using the Energy Reservoir Model (ERM) \citep{vykhodtsev2022review}, which is commonly used in system-level research. The battery’s State of Charge (SoC), denoted by $E[k] \in [\underline{E}, \overline{E}]$, evolves according to the charging ($P^\text{b}_\text{c}$) and discharging ($P^\text{b}_\text{d}$) powers. The ERM model is mathematically represented as follows,
\begin{subequations}
    \begin{align}
    E[k+1] &= E[k] + \frac{\Delta t} {E_\text{C}}\left(\eta^\text{b}P^\text{b}_\text{c}[k] - \frac{1}{\eta^\text{b}}P^\text{b}_\text{d}[k]  \right) \quad \forall k \in K,\label{BM1}\\
         \underline{E} &\leq E[k] \leq \overline{E} \quad \forall k\in K,\label{BM2}\\
        0 &\leq P^\text{b}_\text{c}[k] \leq P^\text{b}_\text{max} \quad \forall k\in K, \label{BM3}\\
        0 &\leq P^\text{b}_\text{d} [k] \leq P^\text{b}_\text{max} \quad \forall k\in K,\label{BM4}\\
        P^\text{b}_\text{c}[k] &= P^\text{b}_\text{max} z[k] \quad \forall k\in K, \label{BM5}\\
        P^\text{b}_\text{d}[k] &= P^\text{b}_\text{max} (1-z[k]) \quad \forall k\in K, \label{BM6}
    \end{align}
\end{subequations}
where $E_\text{c}$ is the energy capacity of the battery in kWh and $\eta^{b}$ is battery efficiency, K is sample size and $\Delta$t is time step between two samples. Additionally, the complementarity constraints in \eqref{BM5} and \eqref{BM6} ensure that the battery cannot be charged and discharged simultaneously. 

\vspace{0.3cm}\noindent\textit{d) Household demand model ($P^\text{d}$)}\\
The household demand, indicated in \eqref{BD}, is modeled as the sum of the base demand ($D$) and the additional power consumption ($P^\text{HP}$) associated with the deployment of controlled loads. 
\begin{equation}
    P^\text{d}_{\tau, h}[k] = D_{\tau, h}[k] + P^\text{HP}_{\tau, h}[k] \quad \forall k \in K, \forall~h \in \mathcal{H}_\tau, \tau \in\text{T},
    \label{BD}
\end{equation}
where K, $\text{T}$, and $\mathcal{H}_\tau$ represent the sets of discrete time instants, transformers in the system, and the houses connected to transformer $\tau$, respectively.
The base demand, $D$, is assumed to be available as data, and the $P^\text{HP}$ is determined through modeling the controlled loads. 
This study presents heat pumps (HPs) as controlled loads, and their model is presented here. However, the same approach can be applied to model air conditioners (ACs) by replacing $Q^\text{a}$ with $-Q^\text{a}$, or adapted for other loads using appropriate models. 

HPs are modeled using the Equivalent Thermal Parameter (ETP) model, which accounts for both house parameters and heat pump energy flow to determine the indoor air temperature~\citep{oyefeso_control_2022}. The governing equations for this model in discrete form are given as:
\begin{subequations}
    \begin{align}
        {T}^\text{a}[k+1] &= \frac{\Delta t}{C^\text{a}} \left( T^\text{m}[k] H^\text{m} - (U^\text{a} + H^\text{m}) T^\text{a}[k] + T^\text{o}[k] U^\text{a}+Q^\text{a} \right), \\\notag
        &= \frac{\Delta t}{C^\text{a}} ( T^\text{m}[k] H^\text{m} - (U^\text{a} + H^\text{m}) T^\text{a}[k] + T^\text{o}[k] U^\text{a} \\\notag
        &~~~~~~~+P^{\text{HP}}[k]  COP[k] + Q^{\text{oil}}[k]), \\
        {T}^\text{m}[k+1] &= \frac{\Delta t}{C^\text{m}} \left( H^\text{m} (T^\text{a}[k] - T^\text{m}[k]) + Q^\text{m} \right).
    \end{align}
\end{subequations}

Here, $T^\text{a}$ represents the indoor air temperature, while $T^\text{m}$ is the inner mass temperature. $C^\text{a}$ is the thermal mass of the air, and $H^\text{m}$ is the conductance between the inner air and solid mass. $U^\text{a}$ represents the building envelope's conductance, and $T^\text{o}$ is the outdoor air temperature. $C^\text{m}$ is the building's thermal mass, and $Q^\text{m}$ is the heat flux to the interior solid mass. The function $Q^\text{m}$ captures the heat flux into the interior mass, which accounts for internal and ambient heat gains and the heating effect from the HP. For simplicity, $Q^\text{m}$ is assumed to be zero in this analysis.

\begin{figure}[htbp]
    \centering
    \includegraphics[width=0.8\linewidth]{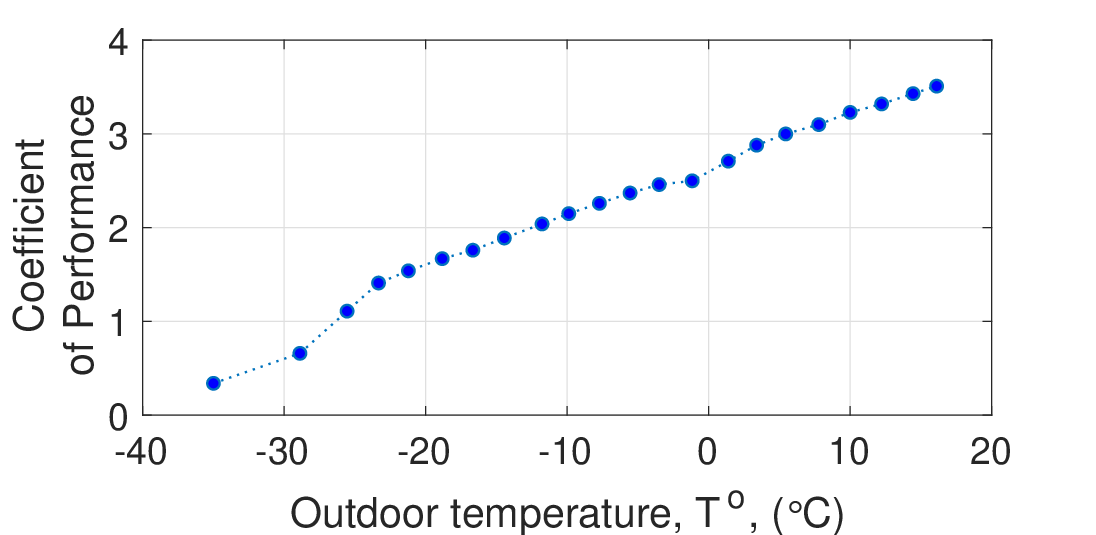}
    \caption{Variation of heat pump's COP w.r.t outdoor temperature \citep{analysisnorth}}
    \label{fig:cop}
\end{figure}

The heat output, $Q^\text{a}$, depends on the outdoor temperature and is given by the addition of thermal energy produced by the heat pump and oil-based heating. The heat pump's thermal energy produced is given by $\text{COP} \times P^\text{HP}$, where $P^\text{HP}$ represents the HP's input power and COP is the coefficient of performance. The COP measures HP's efficiency and is defined as the ratio of thermal energy generated to the electrical energy consumed, which varies with outdoor temperature. Figure~\ref{fig:cop} illustrates the relationship between COP and outdoor temperature observed in the North Arctic region \citep{analysisnorth}. The figure shows that COP decreases significantly at extremely low temperatures, making the heat pump less efficient at the low temperatures.

For extremely cold days, when the outdoor temperature falls below the heat pump's cutoff temperature, $T^\text{o}_\text{cutoff}$, the system switches to oil-based heating. The corresponding heat flux from oil-based heating is denoted as $Q^\text{oil}$. 
This study specifically examines inverter-controlled heat pumps, where $P^\text{HP}$ is treated as a continuous variable. 

\subsubsection{{Optimal Energy Dispatch with Intelligent Coordination}}
\label{ED}
The optimal dispatch algorithm determines the generator ($P^\text{g}$) and battery dispatch ($P^\text{b}$) for a given renewable generation and demand profile. The demand used in the optimal dispatch can either be predefined (base demand) or dynamically determined through the intelligent coordination of energy-efficient appliances. The objective is to minimize the generator cost ($C_{P^\text{g}[k]}$) and the cost of transformer upgrade while satisfying various constraints, as outlined in~\eqref{12}. The cost of transformer upgrade is calculated as $\gamma P^\text{ex}_\tau[k]$, where $\gamma$ is transformer cost per kVA and $P^\text{ex}_\tau[k]$ is the excess power required by the transformer $P_\tau$ due to addition of energy-efficient appliances in the households at a given instant k.

In this formulation, $P^\text{RE}$ and $P^\text{d}$ are input data, while $P^\text{g}$, $P^\text{b}$ and $P^\text{HP}$ are decision variables. The cost coefficient $\alpha$ is derived from the generator cost curve, and the cost coefficient $\gamma$ is obtained from the literature, manufacturer information, or the utility. 

The value of $P^\text{d}$ varies depending on the transition pathway, the size of energy-efficient appliances installed, and their coordination technique. It is calculated as shown in~\eqref{eq:pd}, where $D_\tau$ represents the base demand at transformer $\tau$ (data), and $P^\text{HP}$ denotes the power consumed by the additional controlled load, e.g, a heat pump. Equations~\eqref{eq:constraint3a} and ~\eqref{eq:constraint3b} indicate the ETP model with a heat pump and oil as a heating source. The heating oil considered is stove oil, and $Q_{oil}$ indicated in ~\eqref{eq:constraint3a} is calculated based on the energy density of the stove oil mentioned by EIA. The initial indoor temperature at the beginning of each new day is set to the final indoor temperature from the previous day to maintain continuity in thermal modeling as indicated in \eqref{Tnextday}. 

The problem is formulated as,
\allowdisplaybreaks
\begin{subequations}
\begin{align}
    \min_{P^\text{g}, P^\text{b}_\text{c},P^\text{b}_\text{d},P^\text{HP}} \sum_{k=1}^{K} &(\alpha {P^\text{g}}[k] +C) + \sum_{k=1}^{K} \sum_{\tau=1}^{\mathcal{T}} \gamma P^\tau_\text{ex}[k] , \\ 
\text{subject to:}~~~~~~~~~~~~~~\notag\\\notag
\text{Eq.} ~\eqref{EF1}, ~&\eqref{EF3}-\eqref{EF5}, \text{Eq.}~\eqref{BM1}-\eqref{BM6} \\
P^\text{d}[k]&=\sum_{\tau \in \text{T}} \Big( D_{\tau}[k] + \sum_{h \in \mathcal{H}_\tau} P^{\text{HP}}_{\tau, h}[k] \Big)\label{eq:pd}\\
P_\tau[k] &= D_\tau[k] + \sum_{h=1}^{N_\tau} P^{HP}_{\tau,h}[k] \quad \forall k, \tau \\ P_\tau^{\text{ex}}[k] &\geq P_\tau[k] - P^{\text{max}}_\tau \quad \forall k, \tau \\
 P^\text{b}_\text{c}[k] &\leq P^\text{PV}[k] + P^\text{g}[k] \quad \forall k\in K,\\
E[1] &= E[K+1], \\
  T^\text{a}_{\tau, h}[k+1] = T^\text{a}_{\tau, h}[k]  +& \frac{\Delta t}{C^\text{a}_{\tau,h}} \Big( T^\text{m}_{\tau, h}[k]  H^\text{m}_{\tau,h} - (U^\text{a}_{\tau,h}+ H^\text{m}_{\tau,h})  T^\text{a}_{\tau, h}[k]  \notag \\
      + T_{\text{o}, k}  U^\text{a}_{\tau,h}+& P^{\text{HP}}_{\tau, h}[k]  COP[k] + Q^{\text{oil}}_{\tau, h}[k] \Big), \notag\\
      &\quad \forall k \in K, \forall \tau \in \text{T}, h \in \mathcal{H}_\tau \label{eq:constraint3a} \\ 
    T^\text{m}_{\tau, h}[k+1] = T^\text{m}_{\tau, h}[k] +& \frac{\Delta t}{C^\text{m}_{\tau,h}} \Big( H^\text{m}_{\tau,h}  (T^\text{a}_{\tau, h}[k] - T^\text{m}_{\tau, h,}[k]) \Big), \notag\\
    &\quad \forall k \in K, \forall \tau \in \text{T}, h \in \mathcal{H}_\tau \label{eq:constraint3b} \\
     T^\text{a}_{\tau, h}[1] &= T^\text{a}_{\tau, h}[K+1] \label{Tnextday}\\
    P^{\text{HP}}_{\tau, h}[k] &= 0  \quad \text{if } \min_{t \in \text{day}} \{ T_{\text{o}}(t) \} < T^\text{o}_\text{cutoff}, \notag\\
   Q^{\text{oil}}_{\tau, h}[k] &= 0  \quad \text{if }\min_{t \in \text{day}} \{ T_{\text{o}}(t) \} \ge T^\text{o}_\text{cutoff}\\
     T^\text{a}_{\text{min}} \leq T^\text{a}_{\tau, h}[k] &\leq T^\text{a}_{\text{max}}, \quad \forall k \in K, \forall \tau \in \text{T}, h \in \mathcal{H}_\tau \label{eq:constraint4} \\
    P^\text{HP}_\text{min} \leq P^{\text{HP}}_{\tau, h}[k] &\leq P^\text{HP}_\text{max}, \quad\quad  \forall k \in K, \forall \tau \in \text{T}, h \in \mathcal{H}_\tau \label{eq:constraint5} \\
    0 \leq  Q^{\text{oil}}_{\tau, h}[k] &\leq Q_{\text{max}}^\text{oil}, \quad  \forall k \in K, \forall \tau \in \text{T}, h \in \mathcal{H}_\tau \label{eq:cQoil}
\end{align}
\label{12}
\end{subequations}

\subsection{Criteria Assessment}
\label{criteria_indices}
The generation and consumption data obtained from the system model can then be used to calculate the impact of different transition pathways.
The impact of energy transition pathways can be assessed using four main criteria, viz., technological, economic, environmental, and social. The assessment indices used are indicated in Table~\ref{criteria}. 

\begin{table}[htbp]
    \centering
    \caption{List of criteria across Technological Economic, Environmental, and Social tenets}
    \resizebox{1\linewidth}{!}{
    \begin{tabular}{l l l} \toprule
         \textbf{Criterion} & \textbf{Sub-criterion} & \textbf{Units} \\ \midrule
         \multirow{2}{*}{Technological} & Infrastructure upgrade factor & -- \\
                             & Resource Adequacy & \% \\
                             \hline
         \multirow{5}{*}{Economic}& Add'l infrastructure cost & \$\\
                                & Retail electricity rate & \$/kWh \\
                                & {Customer energy assistance} & \$/kWh \\
                                & Annual electricity cost per house & \$\\
                                & Annual heating (non-electric) cost per house & \$\\
                                & Total energy cost & \$\\\hline
        \multirow{2}{*}{Environmental} & CO\textsubscript{2e} & metric tons\\
                                        & PM\textsubscript{2.5} & \si{\micro\gram}/m\textsuperscript{3}\\\hline
        \multirow{2}{*}{Social} & Energy burden & \$\\
                                & Energy poverty & \$\\ \bottomrule
    \end{tabular}}
    \label{criteria}
\end{table}

\subsubsection{Technological Criteria}
\label{criteria_indices_eco}
The technological assessment considers the impact of intelligent coordination of resources or the addition of renewable resources to the existing system. The criteria considered are the average infrastructure upgrade factor and resource adequacy. 

\vspace{0.3cm}\textit{a) Infrastructure Upgrade Factor (IUF):} This criterion quantifies the extent of infrastructure upgrades required for each proposed transition pathway. The IUF is a normalized metric indicating the average ratio of required upgraded capacity to original capacity across all relevant infrastructure assets within the distribution system.

For each infrastructure resource $r \in \mathcal{R}$ (e.g., transformers, diesel generators, solar PV, wind turbine), it's required upgrade capacity, $C_r^\text{up}$, is first determined. This capacity is calculated based on the maximum of the resource's annual load data, rounded up to the nearest commercially available capacity. The original capacity of the resource is denoted as $C_r^\text{orig}$.  

The overall Infrastructure Upgrade Factor (IUF) for a given transition pathway is then calculated as the average of these capacity ratios across all resources $\mathcal{R}$. 
\begin{align}
    \text{IUF} &= 
    \frac{\sum_{r \in \mathcal{R}} (C_r^\text{up}/C_r^\text{orig})}{|\mathcal{R}|}, 
    \label{eq:average_upgrade_factor}
\end{align}

This metric provides a clear and normalized indication of the infrastructure investment needed to support the changes introduced by a specific transition pathway.

\vspace{0.3cm}\textit{b) Resource Adequacy:}
This criterion evaluates system reliability, \tck{building on the widely-used $N-1$ reliability metric, which ensures the system can operate with the loss of any single resource (e.g., a generator or transmission line). However, the $N-1$ metric does not account for the partial operation of the resource. 
To extend the N-1 metric, we consider measuring the percentage of time an additional resource is required. This provides a more granular reliability metric addressing the partial operation scenarios overlooked by the traditional $N-1$ approach.}



\subsubsection{Economic Criteria}
The economic indices considered are additional infrastructure cost, retail electricity rate, customer energy assistance, annual electric and non-electric heating costs, and the total energy cost.

\textit{a) Additional Infrastructure Cost:} 
This criteria represents the cost incurred due to capacity upgrades necessitated by the transition pathways. These upgrades may include transformer replacements, battery storage expansion, or additional generation capacity required to support the integration of renewable energy sources and electrification measures.

\vspace{0.3cm}\textit{b) Retail Electricity Rate (R):} The retail electricity rate, denoted by R, is determined by the ratio of the total electricity supply cost to the total electricity sold within a given service area. For rural and remote communities, this cost structure often includes generation cost, as there is a higher reliance on localized generation.

\begin{align}
    \text{Electricity rate}(\$/\text{kWh}) &= \frac{\text{Total electricity supply cost}}{\text{Total kWh sold}}
    \label{3}
\end{align}

Generation costs typically comprise both fuel costs (e.g., for diesel or natural gas generators) and non-fuel operational costs (e.g., maintenance, labor, capital recovery for generation assets). 
In many remote and rural communities, the electricity supply is evolving with the integration of Independent Power Producers (IPPs), such as locally owned renewable energy plants (e.g., solar, wind, hydro) and energy storage systems. The inclusion of power purchased from these IPPs significantly impacts the overall electricity rate structure.

The revised formula for determining the electricity rate, accounting for IPP contributions, is
\begin{align} 
R &= \frac{\text{Utility Generation Costs} + \text{IPP Purchase Costs} + \text{O\&M Costs}}{\text{Total kWh sold}},
\notag \\ 
&= \frac{C_\text{fuel} /\eta_\text{fuel}\times E_{\text{fuel-gen}} + C_{\text{O\&M, utility}} + E_{\text{IPP}} \times R_{\text{IPP}} }{\text{Total kWh sold}},
\label{ER_IPP}
\end{align}

Where $C_\text{fuel}$ is the fuel cost per gallon, $ E_{\text{fuel-gen}}$ is the total electricity produced from utility-owned fossil fuel generation, incorporating generator efficiency $\eta_\text{fuel} $(gallons/kWh). $C_{\text{O\&M, utility}}$ accounts for the non-fuel operational and maintenance costs associated with utility-owned electricity production and other utility operations. $E_{\text{IPP}}$ denotes the total electricity (in kWh) purchased from IPPs. $R_{\text{IPP}}$ is the rate paid to IPPs for purchased electricity (e.g., \$/kWh). 


\vspace{0.3cm}\textit{c) Customer Energy Assistance (CEA):} 
Customer Energy Assistance (CEA) mechanisms address the financial burden of electricity expenses by promoting affordability and equitable access to energy. These programs offset costs, whether due to inherently high regional electricity supply costs (e.g., from reliance on costly generation fuels or the small scale and isolation of local grids) or economic vulnerabilities that make standard tariffs unaffordable for certain customer segments. Examples include targeted rate discounts, percentage of income payment plans (PIPPs), and the federal Low Income Home Energy Assistance Program (LIHEAP).

Eligible consumers (e.g., residential, public facilities) receive support for electricity consumption up to a specified limit. The level of assistance provided is generally linked to factors influencing the consumer's energy burden. For programs addressing inherently high regional electricity supply costs, this might involve the underlying cost of electricity provision (e.g., fuel prices, non-fuel operational and maintenance costs, capital recovery for infrastructure, and the cost of local energy sources, including power purchased from independent producers). {For programs addressing economic vulnerabilities, the assistance level is often tied to household income, size, and overall energy expenditure relative to income.}

A common approach compares actual electricity costs to a baseline reference rate ($R_\text{base}$) representing an affordable or target cost, potentially derived from larger grids or policy thresholds. Programs may also use maximum thresholds ($R_\text{max}$) for eligible costs or assistance amounts to ensure fiscal sustainability. The CEA level calculation varies, but generally aims to bridge the gap between actual burden and the baseline. {A general rate differential calculation is presented in \ref{eleRate}.}
\begin{equation}
\text{CEA Level} = \mathcal{f}(\text{R}, R_\text{base}, R_\text{max}, \text{Customer specific factors})
\label{eleRate}
\end{equation}
This function can take various forms, such as proportional benefits scaled by an adjustment factor, direct bill credits, or income-based caps. This approach highlights how diverse support programs address financial disparities in various energy environments.

\vspace{0.3cm}\textit{d) Average annual electricity cost per house:}
The average annual electricity cost per household is determined based on total electricity consumption and applicable cost adjustments. Using the calculated/available consumption data, the total monthly electricity usage for all residential households is used to calculate the electricity rate. From this, the average monthly consumption per household is determined.

In regions where CEA programs are available, a portion of the household's electricity expenses may be subject to cost adjustments or direct support. The impact of CEA on a household's total annual electricity cost will depend on the specific program design. If the CEA program provides a per-unit (e.g., \$/kWh) reduction on a specific volume of consumption (e.g., up to a limit), the total annual electricity cost per household can be calculated as
\begin{align}
   \text{CEA Eligible kWh} \times \text{(R - CEA level)} + \text{non-eligible kWh} \times R. 
   \label{ele_cost} 
\end{align}
{Alternatively, if CEA is provided through other mechanisms (e.g., percentage discounts on the total bill, fixed monthly credits, or income-based payment caps), the annual cost would reflect these adjustments to the standard bill amount. The total annual electricity cost is ultimately the sum of all monthly charges paid by the household after accounting for any CEA benefits.}

\vspace{0.3cm}\textit{e) Average annual heating (non-electric) cost per house:}
Annual non-electric heating costs are determined based on the amount of fuel required for space heating. In Northern climate regions, households rely on heating fuels such as stove oil, propane, or natural gas, with average consumption varying depending on insulation, building efficiency, and climatic conditions~\citep{nwab2022}. The total annual heating cost is calculated as the product of the fuel consumption and the prevailing fuel price per unit. The amount of fuel consumed is calculated using the ETP model \eqref{eq:constraint3a}. 

For transition pathways incorporating heat pumps, non-electric heating costs are adjusted based on the number of extreme cold days when the heat pump is unable to operate efficiently and requires backup heating. This ensures that the analysis accounts for the impact of electrification on household heating expenses while considering temperature-dependent limitations of heat pump operation.

\subsubsection{Environmental Criteria}\label{EnvC}
The environmental assessment focuses on equivalent carbon dioxide (CO\textsubscript{2e}) emissions \eqref{co2} and air quality, specifically PM2.5 levels \eqref{pm25}, using emission factors from the AP42 guidelines. The expressions to calculate these levels are indicated below.
\begin{subequations}
    \begin{align}
    \text{CO}_\text{2e} &= (\text{CO}_2^\text{EI} \times \text{CO}_2^\text{GWP} + \text{CH}_4^\text{EI} \times \text{CH}_4^\text{GWP} + \notag \\ 
     &~~~~~\text{N}_2\text{O}^\text{EI} \times \text{N}_2\text{O}^\text{GWP}) D_\text{g},\label{co2}\\
    \text{PM}_\text{2.5} &= \text{PM}_\text{2.5}^\text{EI} D_\text{g},\label{pm25}
    \end{align}
    \label{Env}
\end{subequations}
where $\text{CO}_2^\text{EI}$, $\text{CH}_4^\text{EI}$, and $\text{N}_2\text{O}^\text{EI}$ are the emission indices (lb/mmBtu) for respective greenhouse gases, $\text{CO}_2^\text{GWP}$, $\text{CH}_4^\text{GWP}$, and $\text{N}_2\text{O}^\text{GWP}$ are their global warming potential values considering 100 years time horizon, and $D_\text{g}$ is the energy generated from diesel (mmBtu). When considering heating fuel, $D_\text{g}$ refers to the heating energy generated in mmBtu, with emission factors provided by AP42 for the specific fuel type.


\begin{table}[h!]
\centering
\caption{Emission index and global warming potential (GWP) values for different greenhouse gases.}
\resizebox{1\linewidth}{!}{\begin{tabular}{lcccccc} \toprule
 Diesel& \multicolumn{3}{c}{Emission Index (EI)} & \multicolumn{3}{c}{Global Warming Potential (GPW)} \\
type & CO\textsubscript{2}  & CH\textsubscript{4} & N\textsubscript{2}O  & CO\textsubscript{2}  & CH\textsubscript{4} & N\textsubscript{2}O\\
& kg/mmBtu &kg/mmBtu & kg/mmBtu & & & \\
\midrule
No.2  & 73.96  & 0.003 & 0.0006& 1 & 28 & 265 \\
No.1 & 73.25 &	0.003	&0.0006&	1&	25	&298\\
\bottomrule
\end{tabular}}
\label{GHG_factors}
\end{table}

\subsubsection{Social Criteria}\label{SocC}
\vspace{0.3cm}\textit{a) Energy Burden: }The indices considered for this criterion are energy burden and energy poverty. Energy burden is calculated as the percentage of the gross household income spent on the energy cost \citep{baker2019}. 
If the individual household incomes are not available, a common practice is to use the median household income for calculating the energy burden \citep{EB_def}. 

\vspace{0.3cm}\textit{b) Energy poverty}: Energy poverty is defined as lack of access to reliable, affordable energy services \citep{baker2019}. It can be assessed through surveys that capture household experiences related to energy affordability, accessibility, and reliability. Surveys typically include questions on heating challenges, such as whether households have gone without heat, struggled to afford heating fuel or electricity, or faced difficulties in obtaining alternative heating sources like firewood. These indicators serve as proxies for understanding the extent of energy poverty in a given community.  
The responses are aggregated to develop an Energy Poverty Index (EPI). The EPI for the existing scenario is calculated as the mean response across selected survey indicators.  
\begin{align}
\text{EPI}_{\text{base}} &= \frac{1}{N} \sum_{i=1}^{N} X_i,
\label{eq:EPI}
\end{align}
where $ N $ represents the total number of survey indicators used to assess energy poverty, and $X_i$ is the fraction of survey respondents observing a specific energy-related hardship (e.g., inability to afford fuel, lack of heating access).  

For transition pathways, the EPI is adjusted based on the reduction in energy costs, reflecting the impact of electrification on affordability. The modified EPI for different TP scenarios is given by
\begin{equation}
\text{EPI}_{\text{TP}} = \text{EPI}_{\text{base}} \times (1 - \text{cost savings factor}_\text{TP}),
\end{equation}
where the cost savings factor represents the percentage reduction in energy expenses relative to the existing scenario. This formulation enables the assessment of how technological interventions, such as heat pumps, contribute to reducing energy poverty by enhancing access to affordable energy. 

\subsection{Multi-criteria Decision Analysis (MCDA)}
\label{MCDA}
An AHP method of MCDA is used to evaluate trade-offs between different transition pathways and to assess how benefits are distributed across scenarios. The impact of each transition pathway is influenced by stakeholder preferences, which determine the weights applied in the weighted sum method. Equation~\eqref{eq9} presents the calculation of the overall index using this approach.  

In this formulation, $\mathcal{S}^{p}$ represents the score for the $p^{\text{th}}$ transition pathway, $c$ denotes the criteria index, $w_c$ is the weight assigned to the $c^{\text{th}}$ criterion, and $\hat{C}_c^{p}$ is the normalized score of each criterion for the considered transition pathway.
\begin{align} \mathcal{S}^{p} = \frac{1}{N}\sum_{c=1}^N w_c \hat{C}_c^{p} \quad \forall p \in \text{TP} \label{eq9} \end{align}
Since the criteria indices ($C_c^{p}$) are measured in different units across various criteria (c.f. Table~\ref{criteria}), they are normalized using min-max scaling~\citep{vafaei2016normalization}, as shown in Equation~\eqref{eq10}. This normalization ensures that all criteria indices are dimensionless and comparable on a standardized scale between 0 and 1.

In \eqref{eq10}, $C_c^{p}$ represents the value of the $c^{\text{th}}$ criterion under the transition pathway $p$. The terms $C_{c,\text{b}}$ and $C_{c,\text{w}}$ denote the best and worst values of the $c^{\text{th}}$ criterion across all transition pathways, respectively. For all criteria except the electricity rate subsidy level, $C_{c,\text{b}}$ corresponds to the minimum value, and $C_{c,\text{w}}$ corresponds to the maximum value. For the electricity rate subsidy level, $C_{c,\text{b}}$ represents the maximum value, while $C_{c,\text{w}}$ represents the minimum value. The absolute value ensures non-negativity, and the result lies within the range $[0,1]$.
\begin{equation} 
\hat{C}{c}^{p} = \left|\frac{C_c^{p}-C{c,w}}{C_{c,b}-C_{c,w}}\right| \quad \forall p \in \text{TP} \label{eq10} 
\end{equation}

\section{Case Study: Renewable Energy Transition Assessment of Shungnak and Kobuk, Alaska}
\label{sec2}
In this section, we apply the multi-criteria assessment framework detailed in the previous section to assess the feasibility and benefits of renewable energy transitions in Shungnak and Kobuk, in Alaska. Shungnak and Kobuk are adjacent Inupiaq tribal communities located north of the Arctic Circle in the Northwest Arctic. Shungnak and Kobuk are approximately 10 miles apart, interconnected via an AC tie line, with the Shungnak microgrid supplying power to Kobuk.
While Kobuk has a separate diesel generator, it is only used in emergencies. Hence, Shungnak and Kobuk largely operate as a single interconnected electrical system. As of 2022, Shungnak and Kobuk comprise 69 and 47 households, respectively~\citep{shungnak, kobuk}.

\noindent\textit{Community-based collaborative assessment:} We collaborated with the Northwest Arctic Borough (NAB) to assess the feasibility of renewable energy transition in the northern, islanded, microgrid communities in the region. With funding from the U.S. DOE, NAB aims to add community solar arrays in all of its eleven communities and heat pumps in all the households in the region to reduce the energy burden in the region. The study partners, including NAB, AVEC, ACEP, Deerstone Consulting, and UVM, met monthly to compile the data, assess, review, and troubleshoot the analysis presented in this and subsequent sections.
\begin{figure*}[htbp!]
    \centering
    \includegraphics[width=0.65\linewidth]{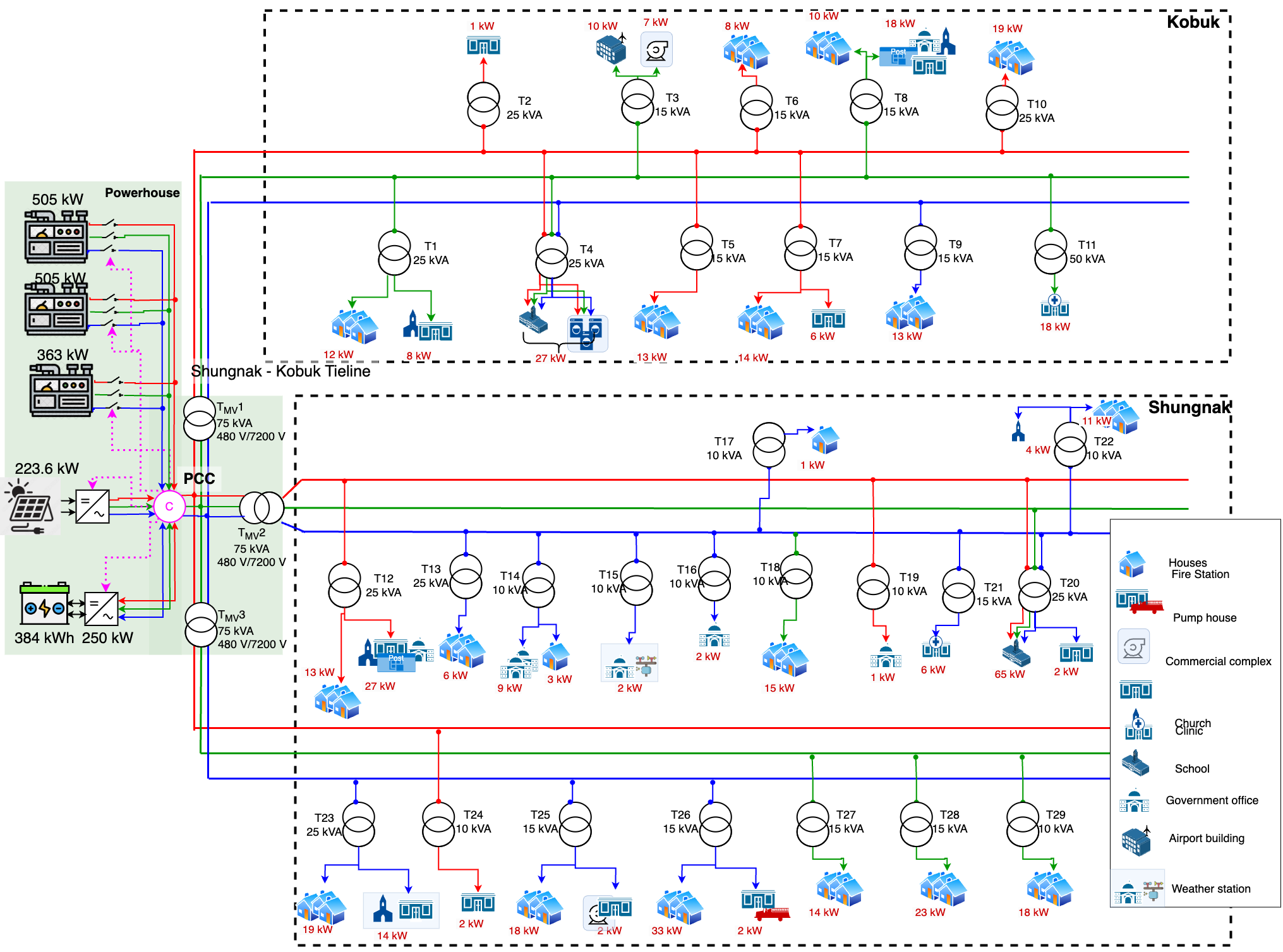}
    \caption{Electrical system block diagram of Shungnak-Kobuk microgrid}
    \label{blockdia}
\end{figure*}

\subsection{Data collection}
The data used for the TEES assessment framework can be divided into five main groups,
\begin{enumerate}
    \item \textbf{Existing energy system infrastructure:} Includes details of the electrical distribution system (Sec~\ref{sec2.1}) such as transformer capacities, diesel generators, solar PV systems, and battery storage. This information was provided by the Alaska Village Electric Cooperative (AVEC), which serves the communities of Shungnak and Kobuk. Relevant data is summarized in Figure~\ref{blockdia} and Tables~\ref{transformer_data} and \ref{HV_tmer}.
    \item \textbf{Time-series operational data:} Actual hourly generation and consumption profiles for diesel generators, PV panels, batteries, and loads for the year 2023 were obtained from the Agito microgrid controller, provided by the Northwest Arctic Borough, a local govt. body. These are shown in Figure~\ref{ED_basecase}.
    \item {\textbf{Sector-wide distribution of demand:}
    Electricity demand for the community is distributed across residential, community (public) and commercial sectors, with residential consumption accounting for approximately 40\%, community (public) facilities for 25\%, and commercial enterprises for 35\%. This distribution is primarily obtained from past customer energy assistance program reports and aligns with findings presented by \cite{devine2004alaska}. }
    \item \textbf{Heat pump and household parameters:} Heat pump specifications are based on manufacturer datasheets and prior literature \citep{oyefeso_control_2022}, with details presented in Tables~\ref{HouseParameters} and \ref{HP_param}. The temperature-dependent coefficient of performance (COP) is sourced from \citep{analysisnorth}, and outdoor temperature data is derived from weather station records. For fuel-based heating, this study assumes 80\% furnace efficiency and stove oil with an energy density to be 138500 btu/gal. 
    \item \textbf{Financial data and assumptions:} Includes historical fuel prices, subsidy values, and economic assumptions provided by the Northwest Arctic Borough. Details are discussed in the economic criteria subsection~\ref{EC_casestudy}.
\end{enumerate} 

\subsection{System Model}
\label{sec2.1}
The electrical system block diagram of the Shungnak-Kobuk microgrid is shown in Figure~\ref{blockdia}. The system is primarily powered by three diesel generators, with capacities of 505 kW, 363 kW, and 505 kW, providing a total generation capacity of 1,373 kW~\citep{shungnak}. Additionally, the system includes 223.6 kW of solar PV capacity \tck{(SolarEdge SE100K)}, and a 384 kWh battery storage \tck{(Blue Planet LX)} with a 250 kW converter system \tck{(EPC PD250)} \citep{shungnak_2021}. These resources are managed by a microgrid controller \tck{(Ageto ARC controller)}, indicated as 'C' in Figure~\ref{blockdia}, which distributes electricity to residential and commercial consumers. In Figure~\ref{blockdia}, the magenta dotted lines represent control lines, while the solid lines denote three-phase power lines. Each phase is indicated by a specific color: red, green, or blue.
\begin{table*}[htbp]
    \centering
    \caption{Rated capacity and maximum load at the LV transformers with residential loads}
    \resizebox{0.95\linewidth}{!}{%
    \begin{tabular}{lcccccccccccccccccccccccccccccc}
    \toprule
    Transformer & T1 & T2 & T3 & T4 & T5 & T6 & T7 & T8 & T9 & T10 & T11& T12 & T13 & T14 & T15 & T16 & T17 & T18 & T19 &T20 &T21 &T22 & T23 & T24 & T25 & T26 & T27 & T28 & T29 \\\midrule
    Capacity (kVA) & 25 & 25 & 15 & 25& 15 & 15 & 15 & 15 & 15 & 25 & 50 & 25 & 25 & 10 & 10 & 10 & 10 & 10 & 10 & 15 & 25 & 10& 25 & 10 & 15 & 15 & 15 & 15 & 10 \\
    \# Houses & 10 & - & - & - & 8 & 6 & 6 & 5 & 5 & 7 & - & 4 & 2 & 1 & - & - & 1 & 6 & - & - & -& 4 & 7 & - & 7 & 14 & 8 & 9 & 6 \\
    Estimated Demand (kW) & 20 & - & - & - & 13 & 8 & 20 & 28 & 13 & 19 & - & 40 & 6 & 12 & - & - & 1 & 15 & - & - & - & 15 & 33 & - & 20 & 35 & 14 & 23 & 18 \\
    \bottomrule
    \end{tabular}%
    }
    \label{transformer_data}
\end{table*}
\begin{table}[htbp!]
    \centering
    \caption{MV transformers rating and maximum load}
    \resizebox{0.4\textwidth}{!}{
    \begin{tabular}{lccc}\toprule
         Transformer & T\textsubscript{MV,1}& T\textsubscript{MV,2}&T\textsubscript{MV,3} \\ \midrule
         Capacity (kVA) & 75 & 75 &75 \\
         \# Houses connected & 47 & 18 & 51 \\
         Total estimated demand (kW) & 89 & 49 & 125 \\\bottomrule
    \end{tabular}}
    \label{HV_tmer}
\end{table}

Power is distributed through three main feeders, one feeding Kobuk and two feeding Shungnak. At the Point of Common Coupling (PCC), each feeder voltage is stepped up using three-phase 422 V/ 7200 V, 75 kVA step-up transformers called medium-voltage (MV) and then distributed through a network of 30 step-down transformers, also called low-voltage (LV) transformers (18 in Shungnak and 12 in Kobuk). Most of the LV transformers are single-phase units, apart from two three-phase transformers installed at schools in Shungnak and Kobuk. The transformers have capacities of 10, 15, or 25 kVA. The red numbers in the figure represent the estimated demand based on the fuse ratings at individual consumer sites. Out of the 30 LV transformers, 19 serve the 116 residential households of Shungnak and Kobuk. The details of the transformer rating and the estimated load for the transformers distributing electricity to the residential customers are indicated in Table~\ref{transformer_data}. Based on the estimated load, five out of the 19 transformers are presently at risk of overload ($\geq$ 1.5 times the rated capacity) if all connected loads are active simultaneously, e.g., T8, T12, T23, T26, and T29, and require an upgrade. As the communities are looking at an alternative to the renewable adoption of heat pumps in all of the households in both these communities, there is a greater risk of overloading of remaining transformers, and upgrades on these transformers are necessary for any additional load. The Ageto microgrid controller records data at the PCC. 

\begin{figure}[htbp!]
    \centering
    \includegraphics[trim = {1.8cm, 0.15cm, 2.5cm, 0.4cm},clip, width=0.95\linewidth]{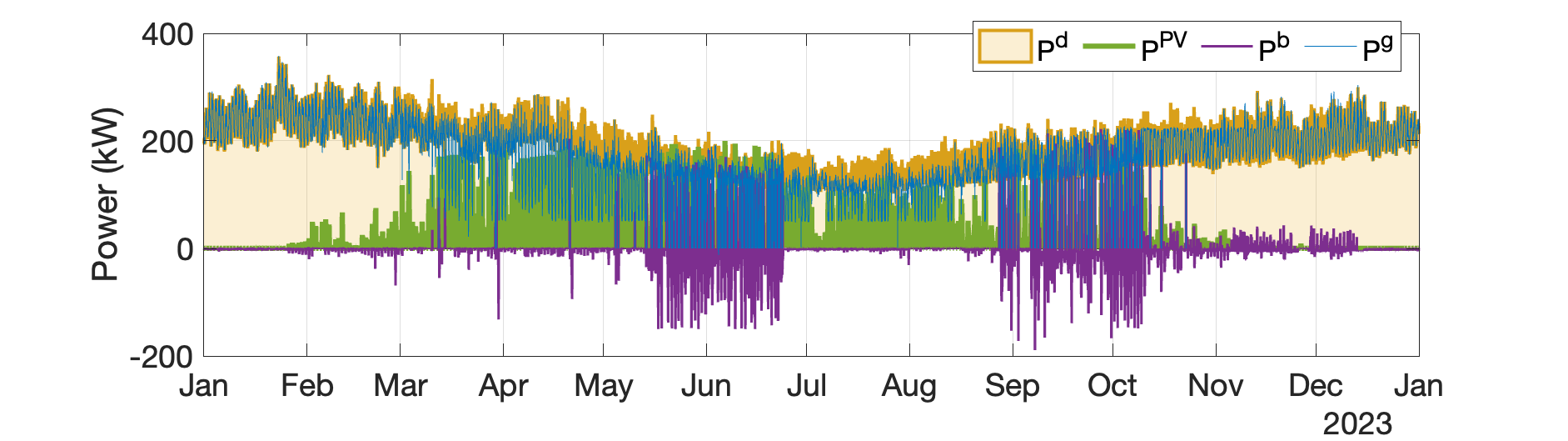}
    \caption{Shungnak-Kobuk grid data recorded by the microgrid controller at the PCC for the year 2023. $P^\text{d}$ represents total demand in kW, $P^\text{g}$ represents power generated by the diesel generator in kW, $P^\text{PV}$ represents solar PV power generation in kW, and $P^\text{b}$ indicates battery power (negative values indicate charging, positive values indicate discharging).}
    \label{ED_basecase}
\end{figure}

Figure~\ref{ED_basecase} presents the historical power values at the Point of Common Coupling (PCC) for the year 2023, detailing total demand ($P^{\text{d}}$), diesel generator output ($P^\text{g}$), solar PV generation ($P^{\text{PV}}$), and battery power ($P^{\text{b}}$), with negative values indicating charging. During winter, minimal PV generation necessitates the diesel generator to cover nearly all demand. Table~\ref{DG} shows the diesel generator operated at 34\% and solar PV at 10\% of their respective annual capacities, indicating significant seasonal solar variability. Notably, summer months depict active battery usage resulting from a temporary manual operational arrangement aimed at increasing generator turn-off time to reduce fuel consumption.

\begin{table}[htbp!]
    \centering
    \caption{Generation details}
    \resizebox{1\linewidth}{!}{\begin{tabular}{l c c c c}
    \hline
        Manufacturer & Model & Capacity & Annual  & Capacity   \\
        & & (kW) & energy (MWh) & factor \\
        \hline
        Diesel generator - Caterpillar & 3456 & 505  & & \\
        Diesel generator - Detroit Diesel & S60 & 363  & 1500 & 34\%  \\
        Diesel generator - Caterpillar & 3456 & 505  & - & - \\
        Solar PV - SolarEdge  & SE100K & 223.6 & 200 & 10\% \\
        \hline
    \end{tabular}}
    \label{DG}
\end{table}
\subsubsection{Modeling of Individual Equipment of the Shungnak-Kobuk System}
\label{sec2.2}
\begin{figure}[htbp!]
    \centering
    \includegraphics[width=0.3\textwidth]{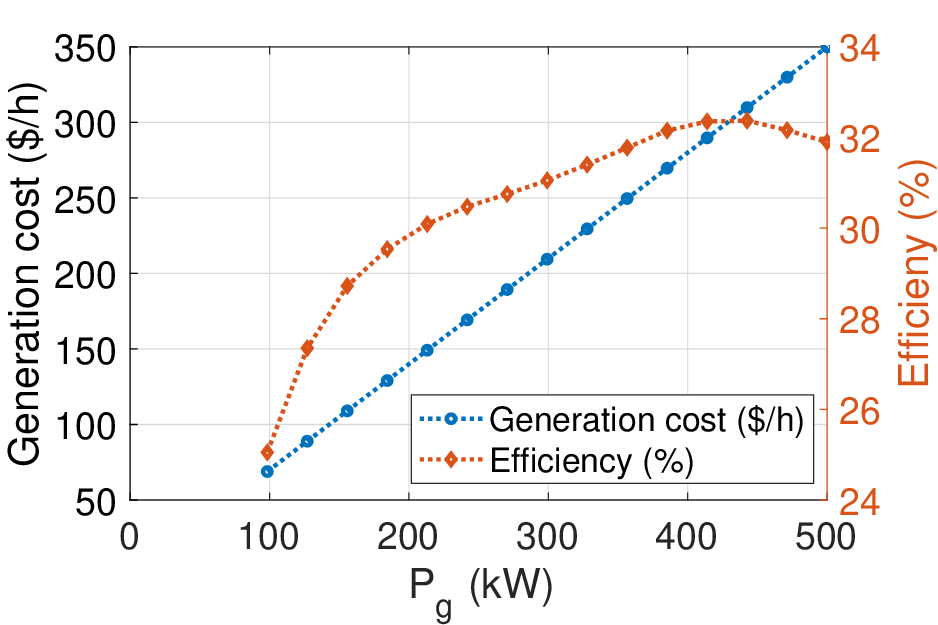}
    \caption{Generator efficiency and cost curve derived from~\cite{nrel_data}.}
    \label{gen_costcurve}
\end{figure}

\noindent\textit{a) Diesel generator model}\\
The generator is modeled as indicated in Section~\ref{sys_model_generic}. 
The fuel cost is considered to be \$10.17 per gallon, as stated in the Alaska Energy Authority's report for Shungnak \citep{pce}. Additionally, the diesel density used is for No. 2 diesel fuel, as outlined by \citet{diesel_density}.
\footnote{The generator also plays a secondary role by supplying waste heat to the water treatment plant. This study does not consider the heat loop, given that increased generator use with added heat pumps would elevate rather than reduce heat output. 
}
The resulting fuel cost curve, along with the NREL generator efficiency curve, is shown in Figure~\ref{gen_costcurve}. This analysis reveals a linear cost curve for the generator represented by \eqref{Cpg_sh}, with $P^g$ in kW. 
\begin{equation}
    C_{P^g} = 0.99P^g + 35.50
    \label{Cpg_sh}
\end{equation}

For this study, the diesel generators are treated as a single entity with a combined capacity equivalent to the total installed capacity. If the required diesel generation capacity exceeds 505 kW, it is assumed that an additional generator must be turned on, potentially reducing the overall reliability of the system. The methodology for calculating the reliability metric is detailed in Section~\ref{criteria_indices_eco}. 
For this case study, the solar PV power output is provided as data and used directly, eliminating the need for a solar PV model.

\noindent\textit{b) Battery model}\\
A battery is modeled as indicated in Section~\ref{sys_model_generic}.
A small 5-minute time step ($\Delta$t = 5 min) is used in this study. This time step is selected because it is fast enough to accurately capture the dynamics of the SoC while maintaining computational efficiency.

\noindent\textit{c) Base transformer demand}\\
To determine the base demand at each specific transformer, the total estimated electricity demand is first disaggregated according to the sectoral percentages (residential - 40\%, community - 25\%, commercial - 35\%). Subsequently, the demand allocated to each sector is then proportionally distributed among the transformers based on the number of service connections associated with that sector at each given transformer.
 
\noindent\textit{d) Heat pump}\\
An HP is modeled as indicated in Section~\ref{sys_model_generic}. The house parameters ($C^\text{a}$, $U^\text{a}$, $H^\text{m}$, and $C^\text{m}$) are adapted from \citet{oyefeso_control_2022} where the authors generated the parameters using GridLAB-D with $\pm 10\%$ random variation for a house size of 2500 sq.ft. The parameters are scaled proportionately to reflect the typical 1000 sq. ft. houses for Shungnak and Kobuk. The range of parameters considered in this study is indicated in table~\ref{HouseParameters}.
\begin{table}
    \centering
        \caption{\tck{Range of house envelop parameters considered \citep{oyefeso_control_2022}}}
    \resizebox{0.2\textwidth}{!}{\begin{tabular}{c c}\toprule
         Parameter & Range  \\\midrule
         $C^\text{a}$ (kWh/$^\circ$C)  & 0.21 - 0.25 \\
         $C^\text{m}$ (kWh/$^\circ$C) & 0.81 - 0.99 \\
         $H^\text{m}$ (kW/$^\circ$C) & 1.76 - 2.17 \\
         $U^\text{a}$ (kW/$^\circ$C) & 0.09 - 0.12 \\\bottomrule
    \end{tabular}}
    \label{HouseParameters}
\end{table}

The Shungnak-Kobuk electrical system is simulated considering optimal energy dispatch (Section~\ref{ED}) and using the diesel generator and battery models. The results obtained are validated with the historical diesel generation and battery power variation data (Figure~\ref{ED_basecase}) for the base case (case with no heat pumps). 

For the optimal energy dispatch $\underline{E}$ and $\overline{E}$ are considered as 20\% and 90\%, respectively, $\eta^\text{b}$ to be 95\%.

The optimal energy dispatch computes diesel generator and battery power profiles for different transition pathways (TP). For each TP, total demand ($P^\text{d}$) is determined based on the heat pump size and the coordination strategies as defined in \eqref{eq:pd} and detailed in Section~\ref{pathways}. 
The total demand profile ($P^\text{d}$) is then used to solve~\eqref{12}, yielding the corresponding diesel generator power ($P^\text{g}$) and battery power ($P^\text{b}$) profiles. Finally, these $P^\text{g}$ and $P^\text{b}$ profiles are utilized to calculate TEES assessment criteria as illustrated in Figure~\ref{analysis_methods}.

\subsection{Transition Pathways}
\label{pathways}
The transition pathways considered in this study are summarized in Table~\ref{pathway}. These communities rely heavily on fuel for space heating and electricity~\citep{nwab2022}. The pathways are designed to evaluate the impact of various strategies that integrate heat pumps (HPs) of different capacities (18 MBtu/h and 12 MBtu/h) with and without intelligent coordination, as well as renewable energy expansion as an alternative to reduce their fuel dependency. The objective is to assess the TEES benefits of these approaches and identify the most efficient transition pathways for the community.  

Four primary transition pathways are examined, (1) no heat pumps (existing system), (2) uncoordinated heat pumps, (3) coordinated heat pumps, and (4) heat pumps with expanded community solar PV. 
For each transition pathway, the annual electricity generation and consumption are calculated. These results are then used to evaluate TEES performance indices and ultimately to determine the transition score. 
\begin{table}[htbp!]
    \centering
    \caption{Overview of transition pathways considered}
    \resizebox{0.4\textwidth}{!}{\begin{tabular}{p{7cm}}
    \toprule
    \textbf{Transition Pathways (TP)} \\\midrule
    1. No heat pumps (existing system) \\
   2. Uncoordinated heat pumps \\
    \hspace{0.5cm}(a) 18 MBtu/h \\
    \hspace{0.5cm}(b) 12 MBtu/h \\
   3. Coordinated heat pumps\\
     \hspace{0.5cm}(a) 18 MBtu/h \\
     \hspace{0.5cm}(b) 12 MBtu/h \\
    4. Heat pumps with expanded community solar\\
     \hspace{0.5cm}(a) 18 MBtu/h coordinated HPs\\
     \hspace{0.5cm}(b) 12 MBtu/h coordinated HPs\\
    \bottomrule
    \end{tabular}}
    \vspace{-0.25cm}
    \label{pathway}
\end{table}

\begin{enumerate}[label={}, leftmargin=1em]
\item \textbf{TP1: No Heat Pumps (base case, existing system)}:
This pathway represents the current state of the Shungnak–Kobuk energy system, where no heat pumps are installed in households. Space heating is primarily provided through conventional methods such as furnaces, wood stoves, or Toyo stoves~\citep{nwab2022}. {The generator and battery dispatch is determined using Equation~\eqref{12}, with $P^{\text{HP}} = 0$ since there are no heat pumps in this scenario, only the baseline electric demand is considered. The TEES performance indices are calculated based on the resulting system operation.} 

\item \textbf{TP2: Uncoordinated Heat Pumps}: 
This scenario builds on TP1 by adding heat pumps. It models the installation of heat pumps in every household in Shungnak and Kobuk, leveraging federal funding support \citep{ADN2024HeatPumps}. The impact of two heat pump capacities, 12 MBtu/h and 18 MBtu/h, on system performance is explored in the study. The plan is to retrofit 100\% of homes with heat pumps, supported by the high uptake observed in other communities in the region, such as Ambler in 2019 \citep{ambler}, where similar federal funding initiatives were available.
Heat pump modeling details are in Section~\ref{sec2.2}. 
This uncoordinated HP scenario uses the same dispatch formulation as Equation~\eqref{12}, with $\gamma = 0$ and $P^{\text{HP}} \geq 0$ to account for the additional load introduced by the heat pumps.

\item \textbf{TP3: Coordinated Heat Pumps}:
This scenario builds on TP2 by introducing coordination among household heat pumps to reduce transformer loading. It is implemented by setting $\gamma = 80~\$/\text{kVA}$ in the objective function of Equation~\eqref{12}, thereby prioritizing load balancing at the transformer level.

\item \textbf{TP4: Heat Pumps with Expanded Solar PV}:
This pathway extends TP3 by adding 250~kW of additional community-scale solar PV to the existing Shungnak–Kobuk system, thereby increasing renewable energy penetration and reducing generator dependency.
\end{enumerate}

\subsection{Criteria Assessment}
\label{criteria_indices_shungnak}
This section discusses the application of assessment criteria from Section~\ref{criteria_indices} to the communities of Shungnak and Kobuk.

\subsubsection{Technological Criteria}
\noindent\textit{a) Infrastructure upgrade factor}: The infrastructure upgrade factor is calculated as specified in Section~\ref{criteria_indices_eco}. For the criteria with HPs, the load on the transformer will increase, requiring upgrades to the transformer. 
For each transformer, the required upgrade capacity $C_i^\text{up}$ is determined based on the maximum of its annual load, rounded up to the nearest commercially available transformer capacity. \tck{The annual transformer load for the year 2023 is obtained by solving \eqref{12} for each transition pathway. Specifically, $P_\tau[k]$ represents the minimum load on transformer $\tau$ in each 5-minute interval for the coordinated HP approach, while, `$D_{\tau}[k] + \sum_{h \in \mathcal{H}\tau} P^{\text{HP}}_{\tau, h}[k]$' represents the transformer load 5-minute interval. The upgrade factor is then calculated as shown in \eqref{eq:average_upgrade_factor}, where $C_i^\text{up}$ and $C_i^\text{orig}$ are the upgrade and original capacities in kVA of the $\tau^\text{th}$ transformer.}

For the base case (TP1), the transformer upgrade factor for each transformer is 1, resulting in an average transformer upgrade factor of 1.

\vspace{0.3cm}\noindent\textit{b) Resource Adequacy:}
Resource adequacy is evaluated following Section~\ref{criteria_indices_eco}. In the base case, the single existing generator meets all demand, resulting in 0\% additional generator runtime. However, increased demand from heat pump operation, influenced by their size and control strategy, may necessitate an additional generator if the current demand exceeds 80\% of the rated capacity of the primary generator. Resource adequacy is then quantified as the percentage of total time this additional generator is online. 


\subsubsection{Economic Criteria}
\label{EC_casestudy}
\noindent\textit{a) Additional infrastructure cost:} For TP1 (existing system), the total cost for the Shungnak-Kobuk system was \$1.3 million \citep{pce}. Any additional infrastructure cost, e.g. transformer upgrades, solar/battery installations, due to different transition pathways is considered in the sub-criterion. The transformer cost is calculated for each transition pathway based on the required upgrade capacity. Table~\ref{tmer_cost} indicates the commercially available transformer sizes and the assumed associated cost \citep{jeremy_how_2024, larson_nodate}. 

\begin{table}[htbp!]
\centering
\caption{Cost of transformers by capacity \citep{jeremy_how_2024, larson_nodate}}
\resizebox{0.4\textwidth}{!}{\begin{tabular}{llll}
\toprule
\textbf{LV transformer} & \textbf{Cost (\$)} & \textbf{MV transformer} & \textbf{Cost (\$)}\\
\textbf{Capacity (kVA)} & &\textbf{Capacity (kVA)} & \\
\midrule
25  & 2,500 & 112.5 & 33,226\\
37.5 & 3,000 & 150 & 36,672\\
50  & 4,000  &225 & 48,259\\
75  & 6,000 & 300& 52,533\\
100 & 10,000 & 500 &70,905\\
\bottomrule
\vspace{-1cm}
\end{tabular}}
\label{tmer_cost}
\end{table}

\vspace{0.3cm}\noindent\textit{b) Electricity rate:} The Shungnak-Kobuk microgrid is operated by the Alaska Village Electric Cooperative (AVEC). Electricity rates are determined as shown in~\eqref{ER_IPP}. Historical data indicated that non-fuel costs have remained relatively stable over time~\citep{shungnak}.  

\vspace{0.3cm}\noindent\textit{c) Customer Energy Assistance:} 
In remote Alaskan communities, the Power Cost Equalization (PCE) program provides customer energy assistance. The PCE program, managed by the Alaska Energy Authority (AEA), helps mitigate high electricity costs~\citep{AEA_pce_webpage}. Eligible residential and community facility customers receive the subsidized rate ($R_{s}$) for up to 750 kWh per month. The difference $R_\text{PCE}$ is paid by the state to the utility.  

The PCE level is calculated by modifying~\eqref{eleRate} as \eqref{pce_level}.
\begin{align}
    R_\text{PCE} &= (min(R,~R_\text{max}) - R_\text{base}) \times \eta_\text{PCE} \label{pce_level},\\
    R_{s} &= R - R_\text{PCE}
\end{align}
where $R_\text{base}$ (specified by AEA) is 19.58 ¢/kWh, $R_\text{max}$ is \$1/kWh, and the electricity distribution efficiency $\eta_{ERA}$ is 95\%~\citep{pce}.
For example, if the electricity rate, R, is 1.2 \$/kWh, the $R_\text{PCE}$ is,
$
(\min(1.2,1) - 0.1985) \times 0.95 = 0.7614 \text{ \$/kWh}.
$
Thus, the consumer would pay $R_{s} =$ \$0.4386/kWh for the subsidized portion of their consumption, while the remaining \$0.7614/kWh would be covered under PCE.

\vspace{0.3cm}\noindent\textit{d) Average annual electricity cost per house:}
Using the 2023 data, the total monthly electricity consumption (kWh) for all 116 residential households, and the average annual electric cost were calculated for TP1 (existing system). The demand is obtained for other TPs using the optimization given in~\eqref{eq:pd}. From this, the average monthly consumption per household is determined. Subtracting 750 kWh from the average monthly consumption provides the amount of PCE-eligible electricity per household. The total annual electricity cost for each household is then calculated using \eqref{ele_cost}.

\vspace{0.3cm}\noindent\textit{e) Average annual heating (non-electric) cost per house:}
Annual non-electric heating costs are determined based on the fuel required for heating. In Shungnak-Kobuk, the average heating fuel consumption is 650 gallons per year \citep{nwab2022}. A fuel cost of \$16.14 per gallon results in an average annual heating cost of \$10,491 per household. Nonelectric yearly heating costs for transition pathways considering heat pumps are calculated based on the number of extreme cold days when the heat pump is turned off.

\subsubsection{Environmental Criteria}
The environmental criteria indices are calculated as described in Section~\ref{EnvC}.
In the Shungnak-Kobuk system, diesel generators use No. 2 fuel, while heating systems rely on No. 1 fuel. 
The $\text{PM}_\text{2.5}^\text{EI}$ factor considered is 5.9 g/MMBtu~\citep{ap42}. Air volume for emissions calculations is based on the combined area of Shungnak and Kobuk (43 km\textsuperscript{2}) with an assumed height of 100 meters.  

\subsubsection{Social Criteria}
The social criteria indices are calculated as described in Section~\ref{SocC}.

Energy poverty is assessed using data from a home heating survey conducted in January 2022 by the NANA Regional Corporation, in collaboration with the McKinley Research Group (Table~\ref{HeatingSurvey}). The survey included 839 respondents from the Northwest Arctic Borough (NAB) region, with 28 respondents from Shungnak (10\% of the population) and 26 respondents from Kobuk (15\% of the population).  

\begin{table}[!t]
\centering
\caption{Survey results used for energy poverty calculations}
\resizebox{0.4\textwidth}{!}{\begin{tabular}{p{5.7cm}p{0.7cm}p{0.7cm}}
\toprule
\textbf{Category} & \textbf{Shun-gnak} & \textbf{Kobuk} \\
\midrule
Sample size (of 839 total) & 3\% & 3\% \\
Without heat in last six months & 43\% & 42\% \\
Home heating issues in last six months &  &  \\
\quad Broken or malfunctioning furnace & 57\% & 70\% \\
\quad Cannot afford heating fuel, electricity & 43\% & 30\% \\
\quad Cannot gather or chop firewood & 61\% & 55\% \\
\quad No access to firewood & 26\% & 20\% \\
\quad Other heating issues & 17\% & 25\% \\
Heating issues impacting water/sewer services & 57\% & 56\% \\
Ability to heat water at home & 75\% & 96\% \\
\bottomrule
\end{tabular}}
\vspace{-0.5cm}
\label{HeatingSurvey}
\end{table}
Survey data were used to assess energy poverty, based on indicators such as heating interruptions, inability to afford heating fuel, and access to alternative heating sources. The Energy Poverty Index (EPI) is calculated using~\eqref{eq:EPI}.
\begin{figure*}[htbp]
    \centering
    \subfloat[Representative winter day]{\includegraphics[width=0.4\linewidth]{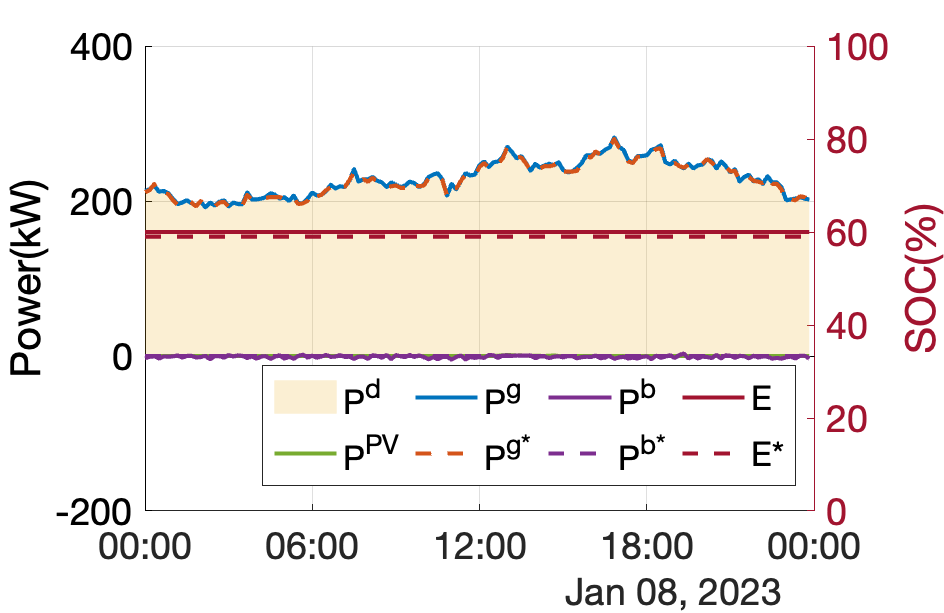}    \label{ED_winter_day}}
     \subfloat[Representative summer day]{\includegraphics[width=0.4\linewidth]{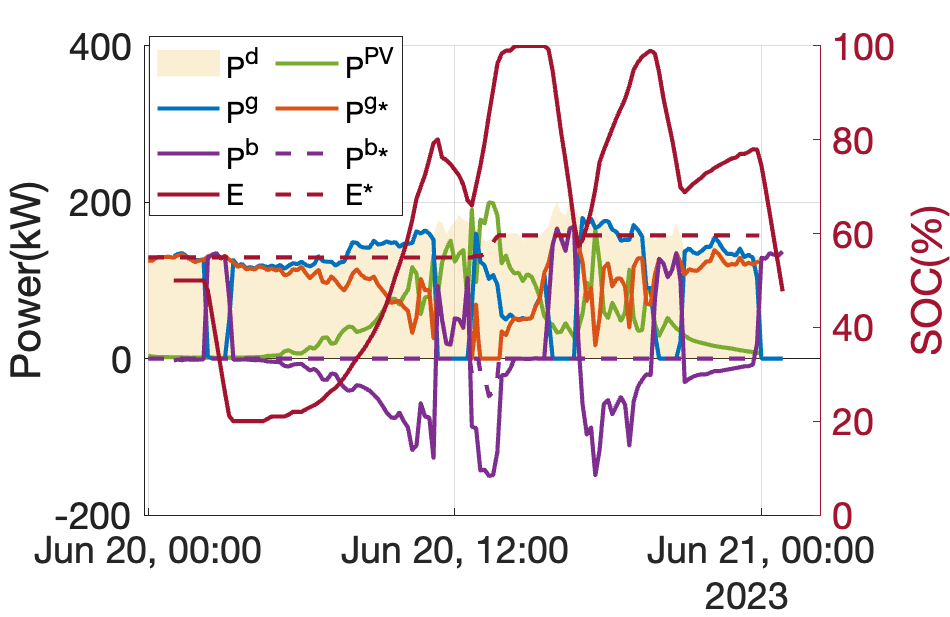}
    \label{ED_summer_day}}
    \caption{Validation of the economic dispatch algorithm with the base case. \tck{Solid lines indicate historical data, whereas dashed lines indicate optimized values.}}
    \label{ED_baseCase_repDay}
\end{figure*}

\section{Results}
\label{sec3}
This section presents the variations in power output (from the generator and transformer) and household temperature across different TPs. 
The implications of these variations on the assessment criteria are further analyzed in the Discussion section. 

\subsection{Transition Pathway 1 (No heat pump):}
This transition pathway establishes a baseline by using historical data on total demand ($P^{\text{d}}$), diesel generation ($P^\text{g}$), solar PV generation ($P^{\text{PV}}$), and battery charging and discharging power ($P^{\text{b}}$), as depicted in Figure~\ref{ED_basecase}. The available demand and PV generation data ($P^\text{PV}$ and $P^\text{d}$) are fed into an optimal energy dispatch problem, outlined in problem~\eqref{12}. The resulting optimized power dispatch for the diesel generator ($P^{\text{g*}}$) and battery (charging $P^{\text{b*}}_{\text{c}}$ and discharging $P^{\text{b*}}_{\text{d}}$) are subsequently compared with the historical data.

The problem is formulated as a mixed-integer linear program (MILP) and solved using the Julia \texttt{JuMP} package with the \texttt{Gurobi} solver. The optimization is performed every week and iterated over an entire year, involving 2016 binary variables per week. The solution time for the entire year’s dataset is 23846 milliseconds.

Figure~\ref{ED_baseCase_repDay} illustrates the optimized energy dispatch for representative winter and summer days, alongside their historical counterparts. Variables marked with '*' denote outputs from the optimization algorithm. The results indicate a close alignment between optimal and historical generator dispatch in winter. However, for the representative summer day, optimal $P^{\text{g*}}$ and $P^{\text{b*}}$ show a distinct trend compared to their historical counterparts, with battery charging occurring for a small duration around noon when there is excess PV generation. This divergence, particularly in battery operation, reveals that the system can be operated optimally even without the manual battery control previously employed as a temporary exploratory activity. Additionally, the historical strategy, by potentially implying frequent generator ramping, might incur higher losses in a real system. Despite these differing operational trends, the optimization confirms the system's near-optimal historical performance in terms of overall costs, even with reduced battery cycling.

\begin{figure}[htbp!]
    \centering
    \includegraphics[width=0.35\textwidth]{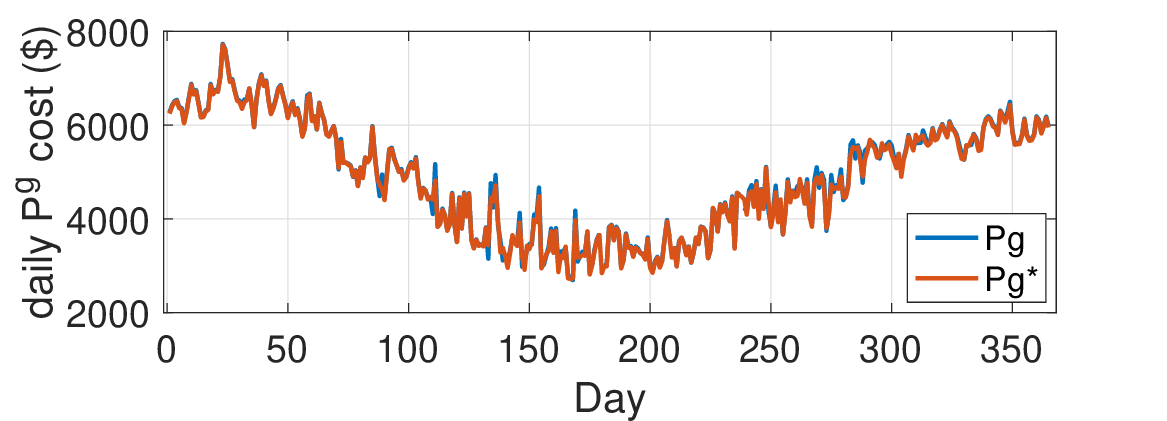}
    \caption{Variation in daily cost of historical value P\textsuperscript{g} and optimized value P\textsuperscript{g*}}
    \label{PgCost_basecase}
\end{figure}

Figure~\ref{PgCost_basecase} highlights the daily cost of $P^{\text{g}}$ and $P^{\text{g*}}$ for the entire year. Across the entire year, the maximum and average daily cost differences between $P^{\text{g}}$ and $P^{\text{g*}}$ are 8\% and 1.1\%, respectively. Notably, the total annual cost of $P^{\text{g*}}$ for 2023 is \$4k (4\%) lower than that of $P^{\text{g}}$. 
A crucial insight from this optimization, however, pertains to battery operation. The optimized solution demonstrates minimal battery charging and discharging activity, contrasting with historical manual management. This is an optimal outcome given the model's economic parameters, which account for battery efficiency losses. For this system, there are no incentives, such as time-varying electricity prices, that would drive frequent battery cycling. Consequently, the model suggests that for minimizing direct generator cost, extensive battery cycling offers limited economic benefit, favoring a largely idle battery over incurring cycling losses.


\subsection{Transition Pathway 2 (Uncoordinated heat pumps):}
For the uncoordinated heat pump scenario, the Shungnak-Kobuk electrical distribution system is simulated with heat pumps installed in each house, as described in Section~\ref{pathways}. The heat pump parameters are provided in Table~\ref{HP_param}, with the HP cutoff temperature ($T^\text{o}_\text{cutoff}$) selected to be the rated minimum outdoor operating temperature listed in the table. This study incorporates a variable coefficient of performance (COP) that depends on outdoor temperature. COP values indicated in figure~\ref{fig:cop} are utilized, as they provide a realistic representation of performance under local weather conditions, unlike manufacturer-provided COP values, which often do not account for such extreme climates.
\begin{figure*}[htbp]
    \centering
    \subfloat[Representative winter day]{\includegraphics[width=0.45\textwidth]{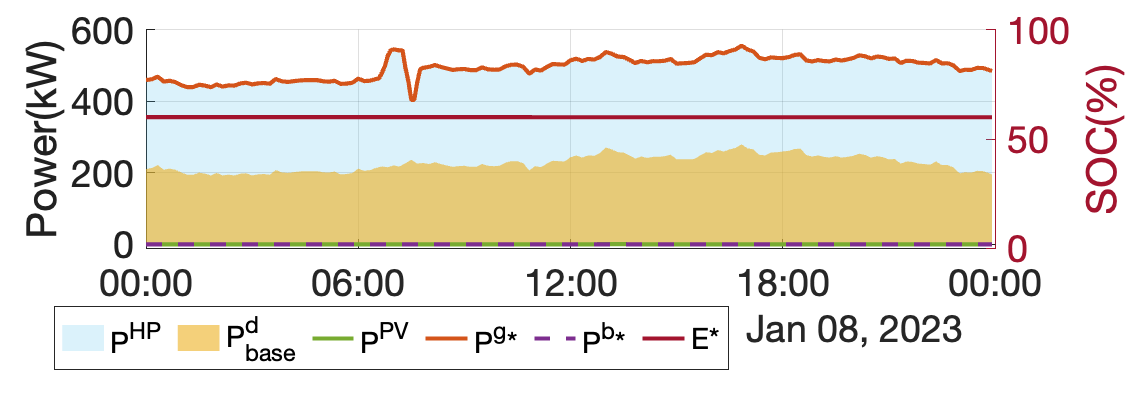}    \label{ED_winter_day_18kdumb}}
     \subfloat[Representative summer day]{\includegraphics[width=0.45\textwidth]{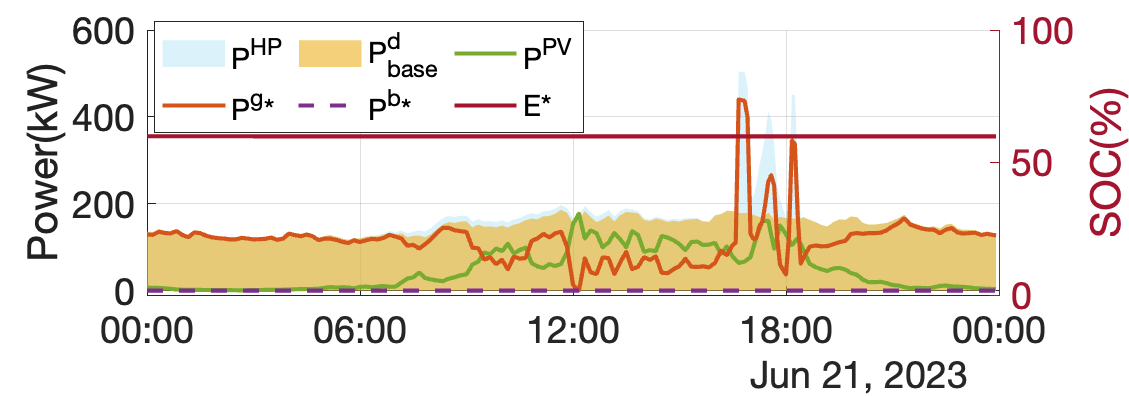}
    \label{ED_summer_day_18kdumb}}\\
    \subfloat[Temperature variation]{\includegraphics[width=0.45\textwidth]{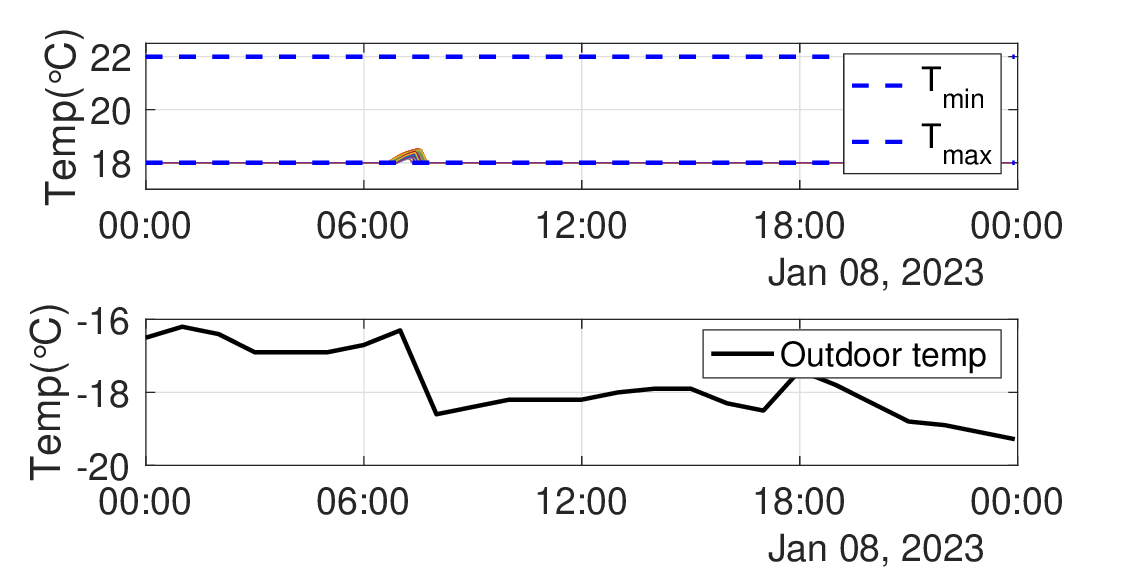}    \label{temp_winter_day_18kdumb}}
     \subfloat[Temperature variation]{\includegraphics[width=0.45\textwidth]{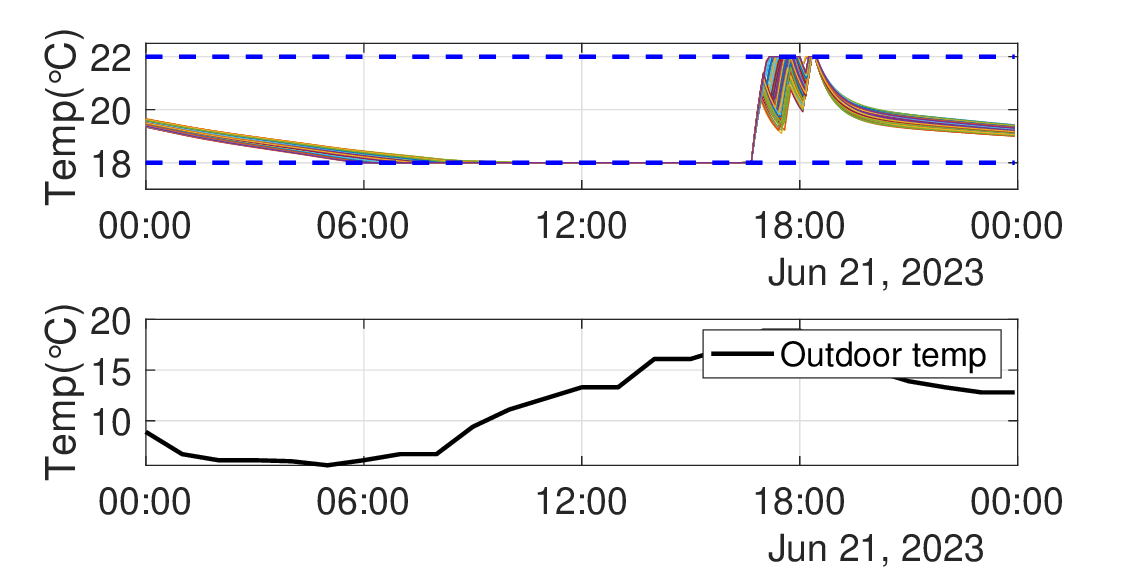}
    \label{Temmp_summer_day_18kdumb}}
    \caption{{Power, indoor ($T^\text{a}$) and outdoor ($T^\text{o}$) temperature variations for representative winter and summer days for 12 MBtu/h uncoordinated HP (TP2b).}}
    \label{ED_12kdumb_repDay}
\end{figure*}
\begin{table}[htb!]
    \centering
    \caption{Heat pump parameters considered for study}
    \resizebox{0.4\textwidth}{!}{\begin{tabular}{p{5cm}cc}\toprule
    Parameter & Value & Value \\\midrule
      Size (MBtu/h)& 18& 12\\
      Rated power (kW)  & 4.14 & 2.81 \\
      {Indoor operating temp ($^\circ$C) }& 18 - 22 & 18 - 22 \\ 
      Heat pump cutoff temperature ($^\circ$C) 
    & {-30} & {-20} \\
    \bottomrule
    \end{tabular}}
    \label{HP_param}
\end{table}

Figure~\ref{ED_12kdumb_repDay} illustrates the power consumption and temperature dynamics for representative winter and summer days under the 12~MBtu/h uncoordinated heat pump (HP) scenario (TP2a). Subfigures (a) and (b) show the power profiles. These include heat pump power consumption ($P^{\mathrm{HP}}$), base demand ($P^{d}_{\mathrm{base}}$), PV generation ($P^{\mathrm{PV}}$), generator output ($P^{g*}$), battery charging/discharging ($P^{b*}$), and the battery state of charge (SOC, plotted on the secondary axis).

During the winter day (Figure~\ref{ED_12kdumb_repDay}~a), demand is primarily driven by space heating, resulting in sustained HP consumption and generator operation, with minimal PV contribution. 
The HP power consumption maintains the indoor temperature at the comfort minimum (Figure~\ref{ED_12kdumb_repDay}~c). A sudden spike in HP power around 6:00 a.m. is observed, due to a sudden reduction in the outdoor temperature, causing HP to draw more power to keep the indoor temperature at Tmin. This peak is also reflected in increased generator dispatch.

In contrast, on the summer day (Figure~\ref{ED_12kdumb_repDay}~b), the outdoor temperatures remain below the minimum setpoint ($T_{\min} = 18^\circ$C) for most of the day. As a result, the HPs continue to operate in heating mode to maintain indoor comfort. A noticeable increase in heat pump (HP) electrical power consumption (Figure~\ref {ED_12kdumb_repDay}~b) is observed around 5:00-6:00 p.m. This surge is directly triggered by indoor air temperatures (Figure~\ref {ED_12kdumb_repDay}~d), which had gradually decreased towards the minimum setpoint throughout the afternoon due to building heat losses, demanding an active heating response. As many homes simultaneously reach this lower heating threshold, their heat pumps switch on, leading to a brief but significant spike in total electrical power demand. Similar studies are conducted for the 18 MBtu/h HP scenario as well; however, the graphical results are not included to avoid repetition. 




\begin{figure*}[t!]
    \centering
    \subfloat[Representative winter day]{\includegraphics[width=0.45\textwidth]{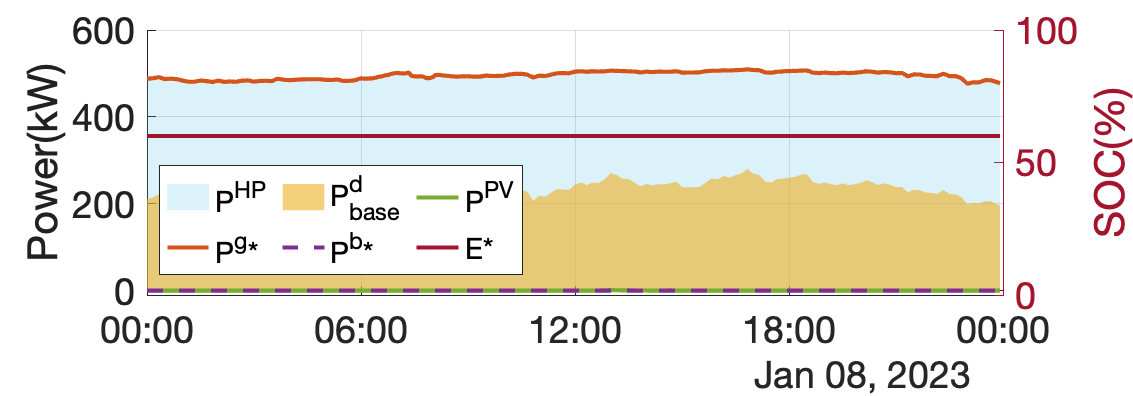}    \label{ED_winter_day_18ksmart}}
     \subfloat[Representative summer day]{\includegraphics[width=0.45\textwidth]{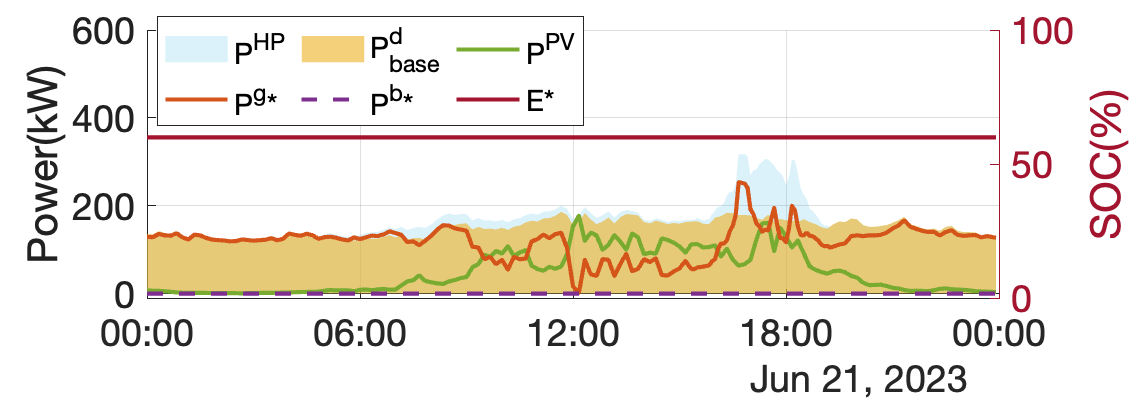}
    \label{ED_summer_day_18ksmart}}\\
    \subfloat[Temperature variation]{\includegraphics[width=0.45\textwidth]{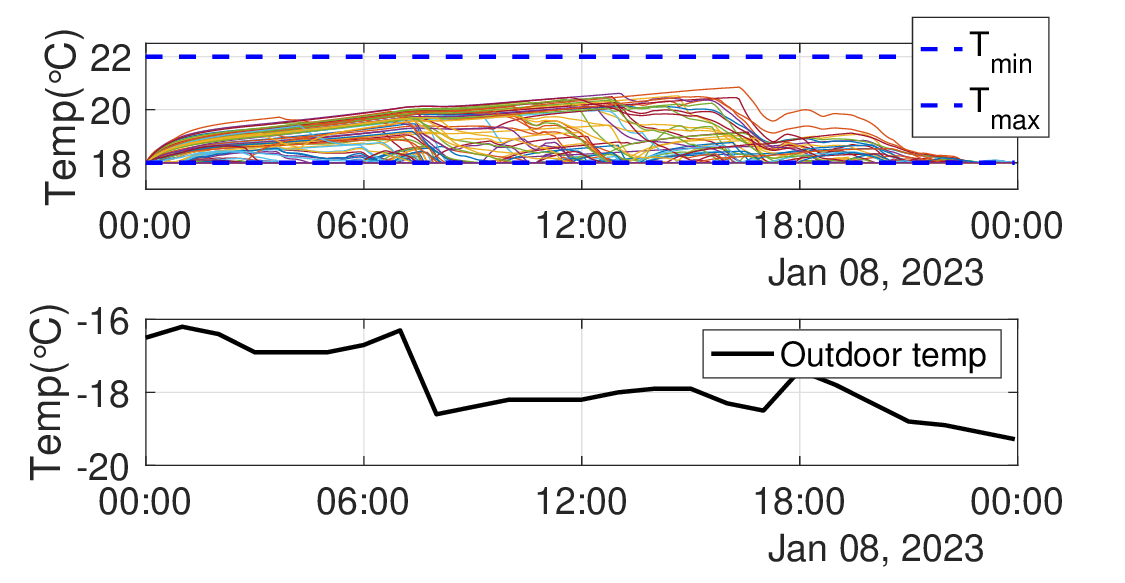}    \label{temp_winter_day_18ksmart}}
     \subfloat[Temperature variation]{\includegraphics[width=0.45\textwidth]{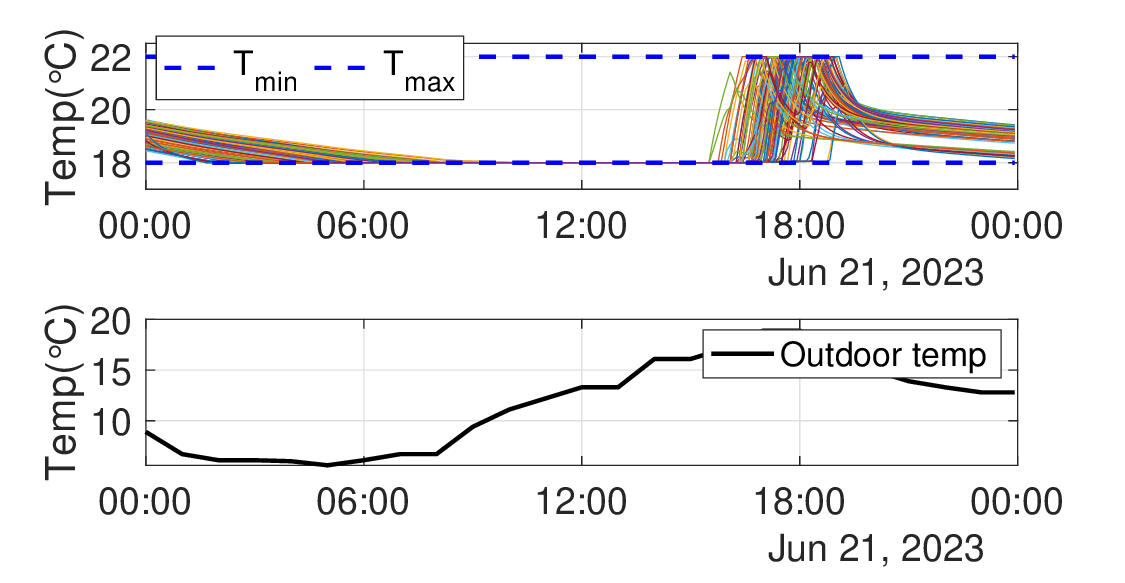}
    \label{Temmp_summer_day_18ksmart}}
    \caption{{Power, indoor ($T^\text{a}$) and outdoor ($T^\text{o}$) temperature variations for representative winter and summer days for 12 MBtu/h coordinated HP (TP3b).}}
    \label{ED_18ksmart_repDay}
\end{figure*}

\subsection{Transition Pathway 3 (Coordinated heat pumps):}
The coordinated HP scenario is also simulated for the entire year. Figure~\ref{ED_18ksmart_repDay} illustrates the variations in power and temperature on representative winter and summer days for the 12 MBtu/h HP system, the cutoff temperature indicated in table~\ref{HP_param}. Unlike the uncoordinated scenario, where HPs operate independently, coordinating HP (Section~\ref{pathways}) across households leads to a flat power profile.

On the representative winter day (Figure~\ref{ED_18ksmart_repDay}a), coordinated dispatch eliminates the peaks seen in the uncoordinated case. Instead, the combined HP + base load tracks a nearly constant value throughout the day. PV generation is negligible, the battery never discharges ($P^{b}=0$), and the SOC sits stably at its target $E^*$ (red line). Indoor temperatures in every house (Fig.~\ref{ED_18ksmart_repDay}c) remain at the $T_{min}$, even as outdoor air swings between –19$^\circ$C and –17$^\circ$C.
On the representative summer day (Figure~\ref{ED_18ksmart_repDay}b), the HP activity remains low, similar to the uncoordinated case, but with notable differences during late afternoon. This event coincides with the presence of solar PV generation (Figure~\ref{ED_18ksmart_repDay}~b), which the optimization algorithm strategically leverages for increased HP consumption to minimize diesel generation costs. The coordination strategy concurrently manages HP power distribution to minimize the overall system peak, thereby directly reducing diesel generation cost.

\subsection{Transition Pathway 4 (Solar PV capacity expansion):}
This section examines the impact of doubling the current solar capacity on various performance criteria, in conjunction with the addition of heat pumps. Both coordinated and uncoordinated HP operation, as well as 18 MBtu/h and 12 MBtu/h HP sizes, are considered. However, results for the coordinated HP scenario only are shown in the table, as uncoordinated HPs yield slightly higher values across all criteria.

\begin{figure}[htbp!]
    \centering
    \includegraphics[width=0.95\linewidth]{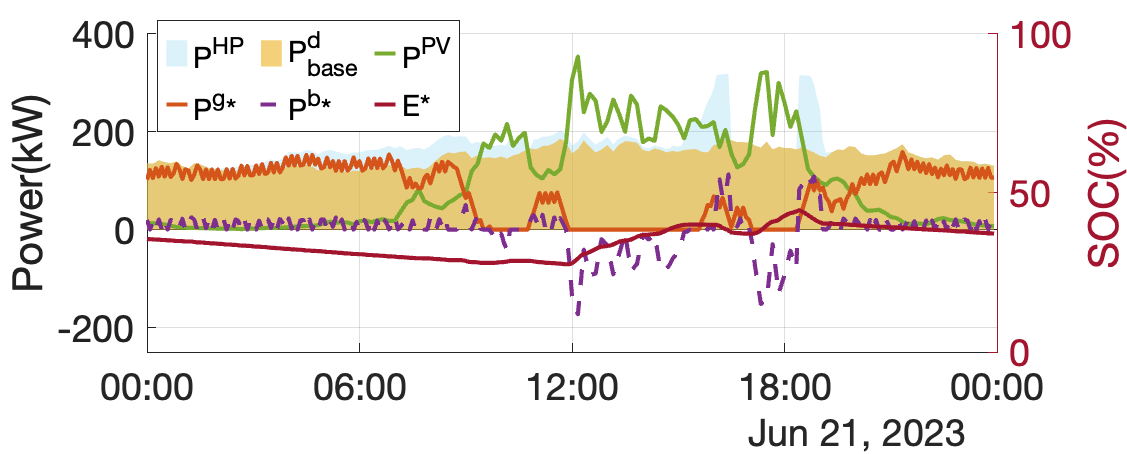}
    \caption{Variation in power during representative summer day for 12 MBtu/h coordinated HP for expanded community solar}
    \label{ED_12k_smart_solar}
\end{figure}

Figure~\ref{ED_12k_smart_solar} illustrates the variations in generation and demand on a representative summer day. During summer, HP usage remains low, resulting in similar demand levels with or without HPs. However, the increase in PV generation capacity necessitates a significantly larger storage capacity, three times the current capacity, to store the additional energy and prevent curtailment. This results in a substantial increase in system cost. In Shugnak-Kobuk, PV expansion incurs no additional costs and was fully funded by the federal government \citep{oced_news}.
In winter, the power demand profile remains unaffected by the expanded solar capacity, as minimal solar generation is available during winter months when HPs are heavily used. Consequently, the addition of solar has a limited impact on winter performance. 

\subsection{Yearly Distribution}
\begin{figure}[htbp!]
\centering
   \subfloat[Integration of 18 MBtu/h HPs]{\includegraphics[trim = {0cm, 0cm, 0cm, 0.5cm},clip, width=0.5\linewidth]{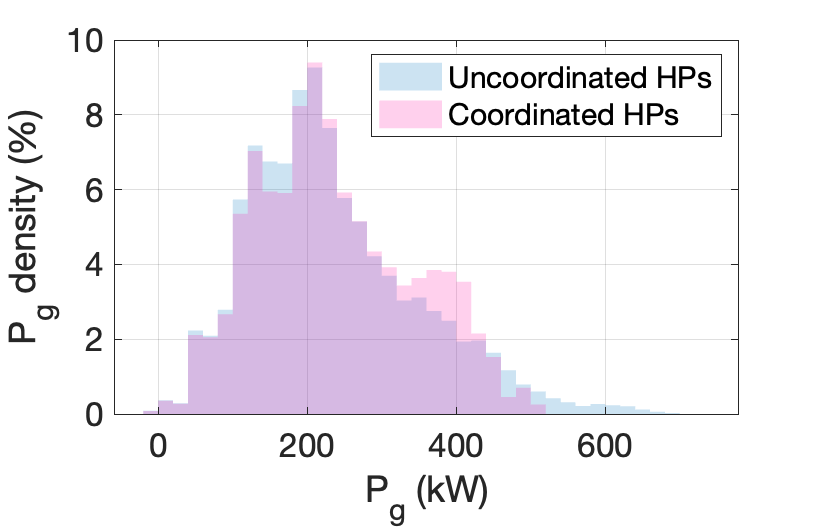}}
    \subfloat[Integration of 12 MBtu/h HPs]{\includegraphics[trim = {0cm, 0cm, 0cm, 0.5cm},clip,width = 0.5\linewidth]{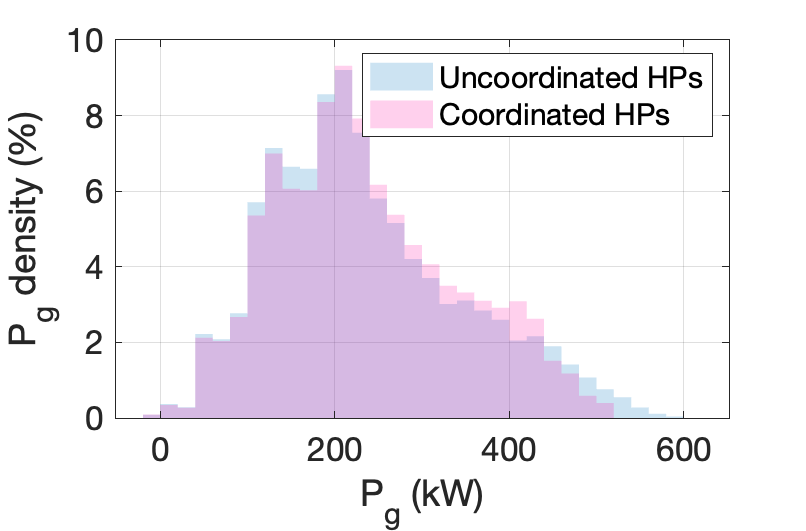}}
    \caption{Distribution of yearly generation by the diesel generator with (a) 18 MBtu/h and (b) 12 MBtu/h integration. HP cutoff temperature, $T^\text{o}_\text{cutoff}$, is considered equal to the rated min outdoor operating temperature (Table~\ref{HP_param})}
    \label{gen_loading}
\end{figure}

Each transition pathway was simulated over a year, and the results are presented in this subsection. Figure~\ref{gen_loading} shows the yearly distribution of diesel generator output for the uncoordinated (TP2) and coordinated (TP3) approaches for (a) 18 MBtu/h and (b) 12 MBtu/h heat pumps, operated with the HP cutoff temperature ($T^\text{o}_\text{cutoff}$) as listed in Table~\ref{HP_param}. Coordinating the operation reduces peak generation demand on the diesel generator, enhancing system reliability by allowing a single 500 kW generator to meet the load while keeping the remaining two gensets available for emergencies. Additionally, the increased demand from heat pumps brings the generator closer to its rated capacity, improving efficiency. The coordinated approach also smooths generation requirements, lowering peak values and distributing the generation more evenly throughout the year.

\begin{figure*}[t!]
    \centering
    \subfloat[Integrating 18 MBtu/h HPs]{\includegraphics[trim = {1cm, 0.5cm, 1cm, 1cm}, width=0.5\textwidth]{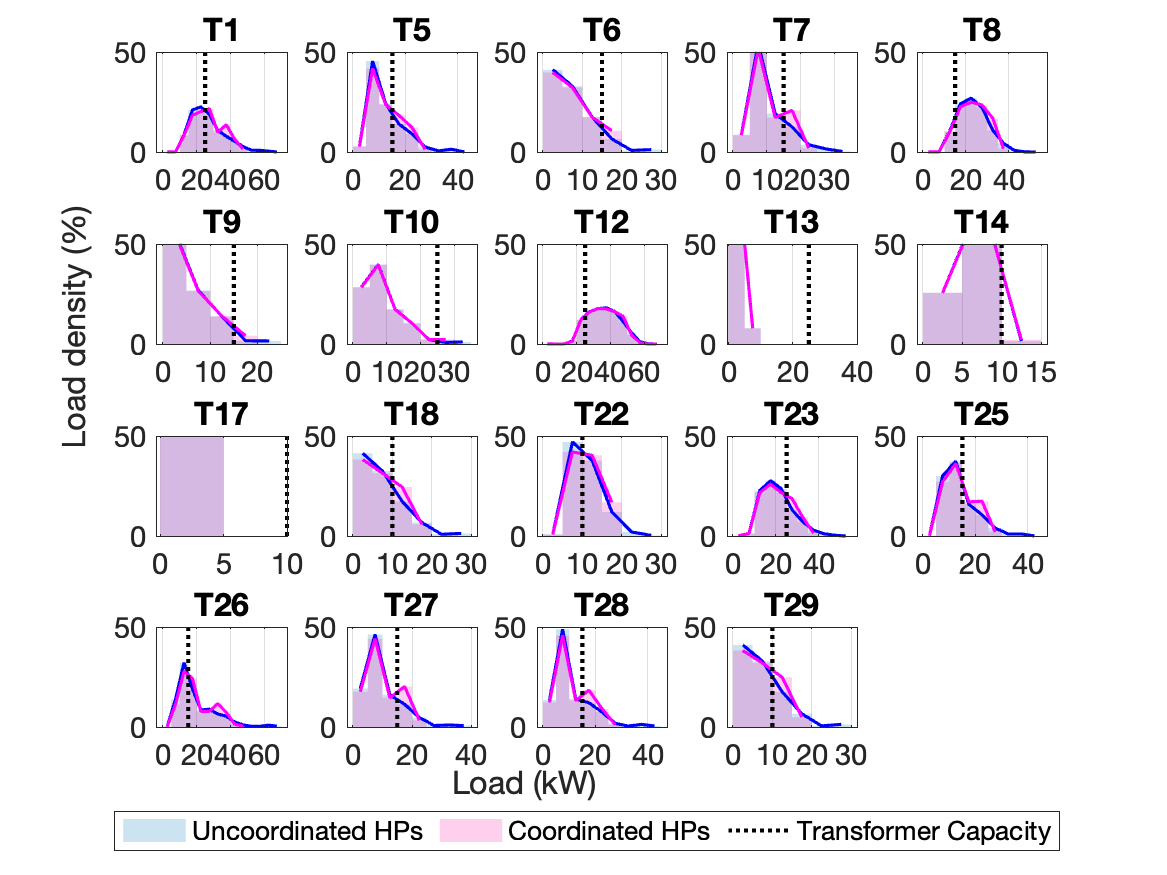}}
    \subfloat[Integrating 12 MBtu/h HPs]{\includegraphics[trim = {1cm, 0.5cm, 1cm, 0.3cm},width = 0.5\textwidth]{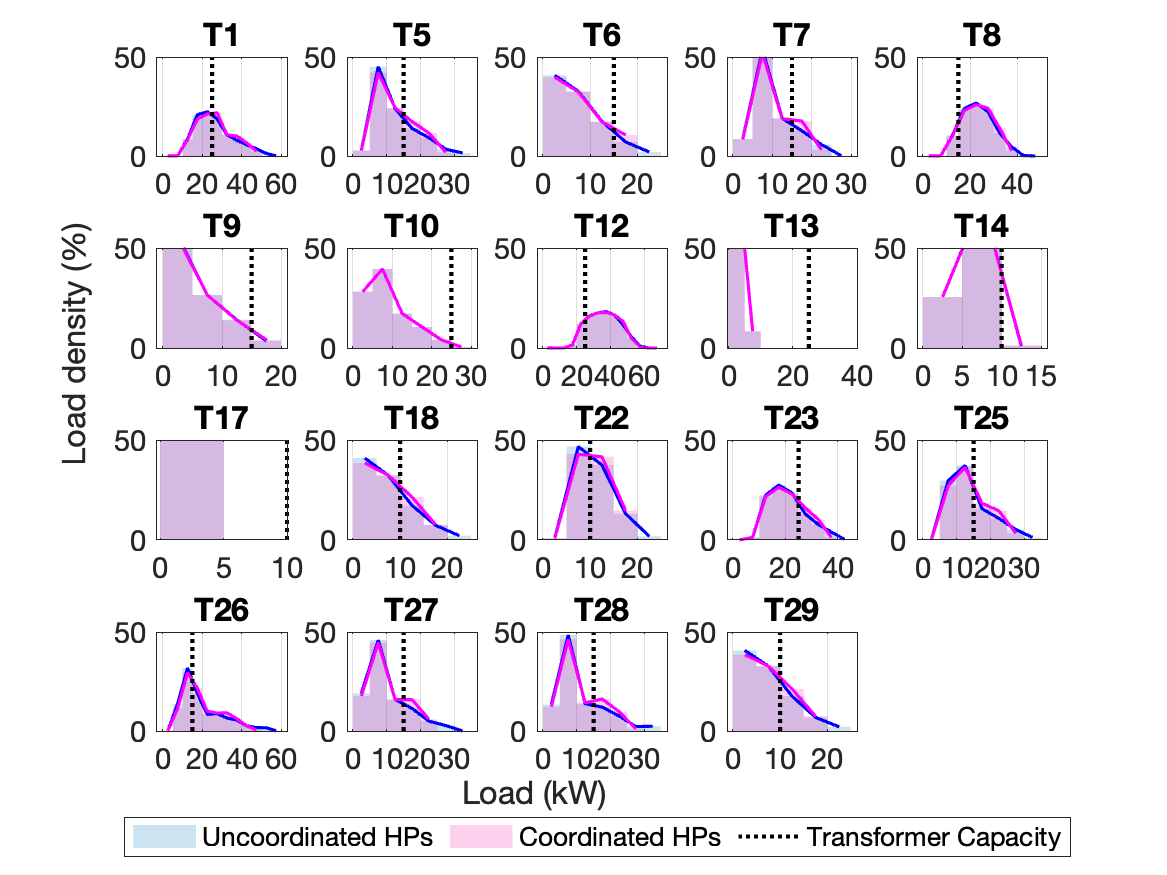}}
    \caption{{Distribution of yearly load on distribution transformers integrating (a) 18 MBtu/h HPs (TP2a and TP3a) and (b) 12 MBtu/h (TP2b and TP3b). Black dotted lines indicate existing transformer capacity. 
    }}
    \label{Tmer_loading}
\end{figure*}

Figure~\ref{Tmer_loading} illustrates the distribution of yearly load on transformers for the 18 MBtu/h and 12 MBtu/h HPs, comparing load density for both control approaches. Each subfigure corresponds to a specific transformer (T1 to T19), with load density represented as a percentage on the y-axis and load in kilowatts (kW) on the x-axis. Transformer capacity limits are indicated by dotted lines.

For the 18 MBtu/h heat pumps (Figure~\ref{Tmer_loading}a), uncoordinated operation exhibits a wider spread in load distribution, frequently reaching or exceeding transformer capacity for several transformers. This indicates a higher risk of overloading transformers under uncoordinated control, necessitating upgrades for 15 out of 19 transformers. In contrast, coordinated HP operation demonstrates a more concentrated load distribution, with load density peaking at lower load levels. Although upgrades are still needed in the coordinated approach, transformer stress is reduced compared to the uncoordinated scenario. Similar results are observed for the 12 MBtu/h heat pumps.

\begin{figure*}[htbp!]
\centering
    \subfloat[]{\includegraphics[width = 0.55\textwidth, trim = {3cm, 0cm, 3.5cm, 0cm}, clip]{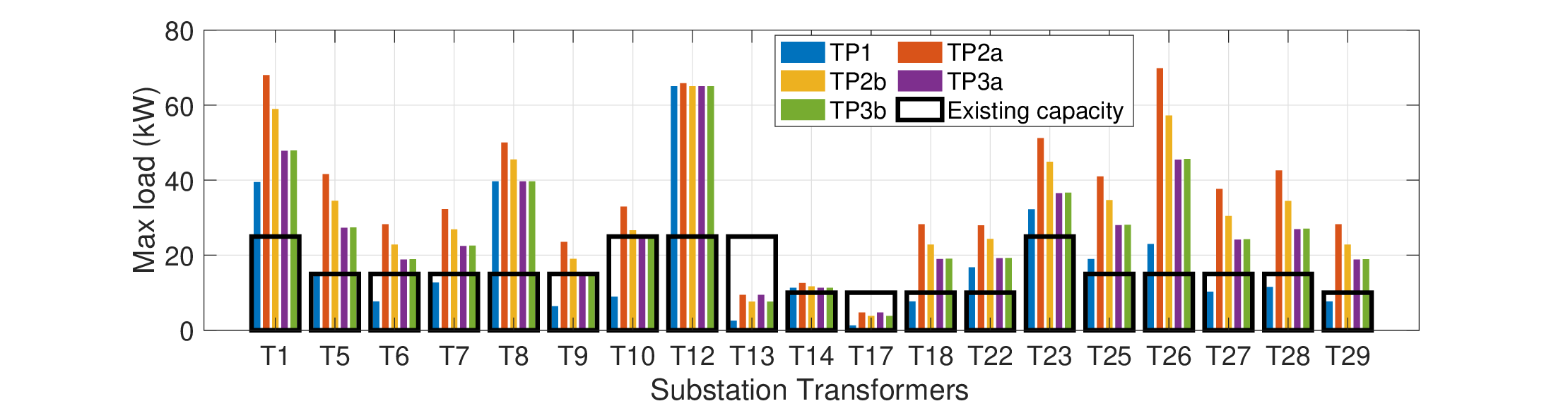}}
    \hspace{1cm}
    \subfloat[]{\includegraphics[trim = {0cm, 0cm, 0.5cm, 0cm}, clip,width = 0.17\textwidth,]{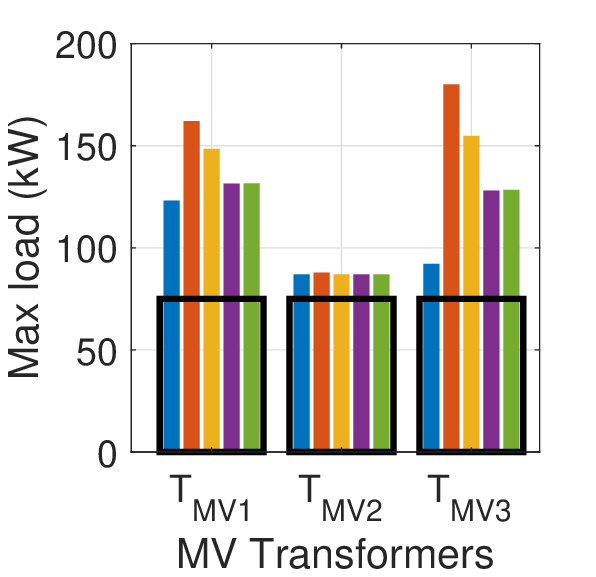}}
    \vspace{-0.3cm}
    \caption{{Technological impact: transformer capacity required for each transition pathway. (a) LV transformers (b) MV transformers.}}
    \label{Tmer_cap_upgrades}
\end{figure*}

Figure~\ref{Tmer_cap_upgrades} illustrates the transformer capacity upgrades required for each transition pathway. For Low Voltage (LV) transformers (Figure~\ref{Tmer_cap_upgrades}~a), TP1, the base case scenario, represents the original transformer capacities derived from solving the economic dispatch for TP1 and considering the base transformer demand as indicated in Section~\ref{sec2.1}. 
In scenarios involving the addition of heat pumps, significant upgrades are required; 15 out of 19 transformers need upgrades for TP2a (uncoordinated 18 MBtu/h HPs), and 14 out of 19 transformers for TP2b through TP3b, indicating that the maximum load on these transformers exceeds 1.5 times their existing rated capacity. TP2a, which involves uncoordinated 18 MBtu/h HPs, necessitates the most substantial upgrades, whereas TP2b, with uncoordinated 12 MBtu/h HPs, requires comparatively lower capacity upgrades. The coordinated approach (TP3a and TP3b) requires similar capacities for both 18 MBtu/h and 12 MBtu/h HP sizes, indicating the smoothing effect of coordination. The solar PV capacity expansion transition pathways (TP4a and TP4b) are not explicitly included in this figure, as increased solar capacity does not affect transformer capacity requirements; their required capacities are therefore assumed to be identical to those in TPs 3a and 3b, respectively.

For Medium Voltage (MV) transformers (Figure~\ref{Tmer_cap_upgrades}~b), the figure indicates the maximum load experienced among all phases for each transformer. Since these are 3-phase transformers, a higher capacity in one phase necessitates upgrading other phase capacities to the same level; therefore, only the phase experiencing the maximum load is shown. Two out of three MV transformers require upgrades. The maximum capacities are necessary for TP2a, while the minimum capacities are needed for TP3a or TP3b.

The analysis over the entire year of 2023 reveals that the coordinated HP approach leads to the smoother \( P^{\text{g}} \) profile. \tck{The generator efficiency also improves slightly compared to the uncoordinated HP approach. The average generator efficiency obtained for TP3 using the generator cost curve and ED solution is 12.23 kWh/gallon and 12.13 kWh/gallon for the 18 MBtu/h HPs and 12 MBtu/h HPs, respectively.} The resulting \( P^{\text{g}}\) values are utilized to calculate various performance indices, demonstrating improvements in system efficiency and reduced operational costs, as shown in Table~\ref{indices}. 

\section{Discussion}
\label{sec4}
This section provides a comparative techno-economic, environmental, and social assessment and the multi-criteria decision-based transition scores for the transition pathways considered for Shugnak and Kobuk, and we talk about how this multi-criteria assessment framework can be scaled to other communities. The yearly generator and battery dispatch data from the previous section are used to evaluate the technological, economic, environmental, and social criteria using the assessment equations outlined in section~\ref{criteria_indices_shungnak}.
\begin{table*}[h!]
\centering
\caption{Technological, Economic, Environmental, and Social criteria assessment for different transition pathways. $T^\text{o}_\text{cutoff} = -20^\circ$C}
\resizebox{\textwidth}{!}{\begin{tabular}{p{4cm} l l l  l l l l }
\toprule
& & \multicolumn{2}{l}{Uncoordinated HPs}& \multicolumn{2}{l}{Coordinated HPs} & \multicolumn{2}{l}{PV capacity expansion} \\
TP & 1 & 2a & 2b & 3a & 3b & 4a & 4b\\
Criterion  & No Hp   & 18 MBtu/h & 12 MBtu/h & 18 MBtu/h & 12 MBtu/h & 18 MBtu/h & 12 MBtu/h\\ 
\midrule
\textbf{Technological} & &    &       &       &  & &      \\ 
Avg. transformer upgrade factor  &  \textbf{1}& 2.4  & 2.0  & 1.7   & 1.7  & 1.7 & 1.7 \\ 
Resource Adequacy(\%)  & \textbf{0} &  2.5  &  1.7  &   \textbf{0}    & \textbf{0} & \textbf{0} & \textbf{0}\\ 
\hline
\textbf{Economic}  &   &   &    &   &  \\ 
Add'l infrastructure cost  & \textbf{0}   & \$172k       & \$158k       & \$148k    &   \$147k     & \$148k    &   \$147k     \\ 
Retail ele. rate (\$/kWh)  & 0.99   & 0.93&    0.93    & 0.93   &0.93 &\textbf{0.91} & \textbf{0.91}   \\ 
PCE level (\$/kWh)  &  \textbf{0.76}  & 0.70   &    0.70   &  0.70  &  0.70 & 0.68 & 0.68\\ 
Subsidized rate (\$/kWh) & 0.23 & 0.23   &    0.23   &  0.23  &  0.23 & 0.23 & 0.23   \\ 
Avg annual electricity \newline \hspace{3mm} cost per household      & \textbf{\$2.40k} & \$4.43k &  \$4.41k      & \$4.41k   &    \$4.40k   & {\$4.40k }  &    {\$4.37k }      \\ 
Avg annual heating cost  \newline \hspace{3mm}(non-electric) per house & \$10.5k  & \textbf{\$4.24k}    &   \textbf{\$2.42k}&  \textbf{\$2.42k}    &  \textbf{\$2.42k } &  \textbf{\$2.42k}    &  \textbf{\$2.42k}     \\ 
Total annual energy cost per household& \$12.9k &  \$8.67k    &     \$8.65k   &  \$8.66k  & \$8.64k & \textbf{\$8.64k} & \textbf{\$8.61k}  \\ 
\% saving       &   --     &  32.7\%  & 32.9\%    & 32.8\%   &   33.0\% &  \textbf{32.0\%}  & \textbf{33.3\% }    \\ 
\midrule
\textbf{Environmental}   & &    &       &      &      \\ 
Annual CO\textsubscript{2e} emission       & &    &    & &  \\ 
\hspace{3mm} From power gen      & 372    &  520  &    518     &  518  &  516  &  \textbf{470}  &  \textbf{470}\\ 
\hspace{3mm} From heating      & 722   &   \textbf{312} &    \textbf{312}   & \textbf{312}  &   \textbf{312} & \textbf{312}  &   \textbf{312}  \\ 
\hspace{3mm} Total reduction    & -- &   23.9\%      &     24.1\%    &   24.1 \% & 24.3\%  &    \textbf{28.5\%} & \textbf{28.5\%}       \\ 
PM2.5 (µg/m3)  & 21.2    &    17.9&     17.9   & 17.8   & 17.8 & \textbf{16.8}   & \textbf{16.8}   \\ 
\midrule
\textbf{Social} & &    &       &      &      \\ 
Energy Burden     &   19.8\%        & 13.3\%   &   13.3\%    & 13.3\% &  13.3\%  &  {13.2\%} &  \textbf{13.1\%} \\ 
Energy Poverty    &  40.0\% & 26.9\% & 26.8\% &  26.8\%  & 26.8\% & {26.7\%} & \textbf{26.7\%} \\ 
\midrule
\textbf{Transition score} & 0.42 & 0.52 & 0.58 & 0.62 & {0.63}  & 0.74 & \textbf{0.75} \\
\bottomrule
\end{tabular}}
\label{indices}
\end{table*}

Table~\ref{indices} presents the assessment results for all considered transition pathways at a heat pump (HP) cutoff temperature of $-20^\circ$C. This cutoff temperature is based on the rated minimum outdoor operating temperature of the 12 MBtu/h HP, and the 18 MBtu/h HP uses the same cutoff for consistency. The analysis includes 18 MBtu/h and 12 MBtu/h HPs under uncoordinated and coordinated operations and scenarios incorporating expanded solar PV capacity.

Technologically, uncoordinated heat pump scenarios (TP2a and TP2b) result in higher transformer upgrade factors (2.4 and 2.0, respectively), while coordinated operation (TP3a, TP3b) and PV expansion (TP4a, TP4b) reduce this requirement to 1.7. Resource adequacy, which indicates the duration of extra resources to be leveraged, has a nonzero value for TP2. This implies that introducing heat pumps without coordination requires supplementary resources to meet the increased demand. Coordinated operation improves system reliability by reducing the need for additional resources.

From an economic perspective, substantial cost savings are observed across all scenarios, including heat pumps. Compared to the base case (TP1), total annual energy costs per household are reduced by up to 33\% for uncoordinated HPs. Coordinated heat pump scenarios (TP3a and TP3b) offer slightly greater savings. The most significant savings are achieved in PV capacity expansion scenarios (TP4b), reaching 33.3\%. These reductions are primarily driven by the shift from diesel-based heating to electricity, which is cheaper per unit. Average non-electric heating costs fall dramatically from \$10.5k annually in the base case to \$4.41k with heat pumps. Electricity rates also drop slightly, from \$0.99/kWh in the base case to \$0.93/kWh with heat pumps. This reduction is because of increased demand for heat pumps. The PCE level shown in the table reflects the state's per-kWh subsidy. The subsidized rate indicates the reduction applied to eligible residential consumption. Although the deployment of heat pumps lowers the overall level of subsidy received, the effective rate paid by eligible households remains unchanged. In contrast, consumers paying the full retail rate (e.g., commercial customers) benefit directly, as heat pump adoption reduces the overall retail rate.

From an environmental perspective, the introduction of heat pumps leads to a significant reduction in CO\textsubscript{2e} emissions. Household fuel-based heating-related emissions decrease from 722 metric tons in the base case to 312 metric tons due to heat pump adoption. Although emissions from power generation increase slightly due to increased electricity usage, the total emission reduces due to a significant reduction caused by household heating. Total emissions reductions reach up to 28\% in the case of a heat pump with solar expansion.

\begin{figure}[t!]
    \centering
    \includegraphics[width=0.9\linewidth]{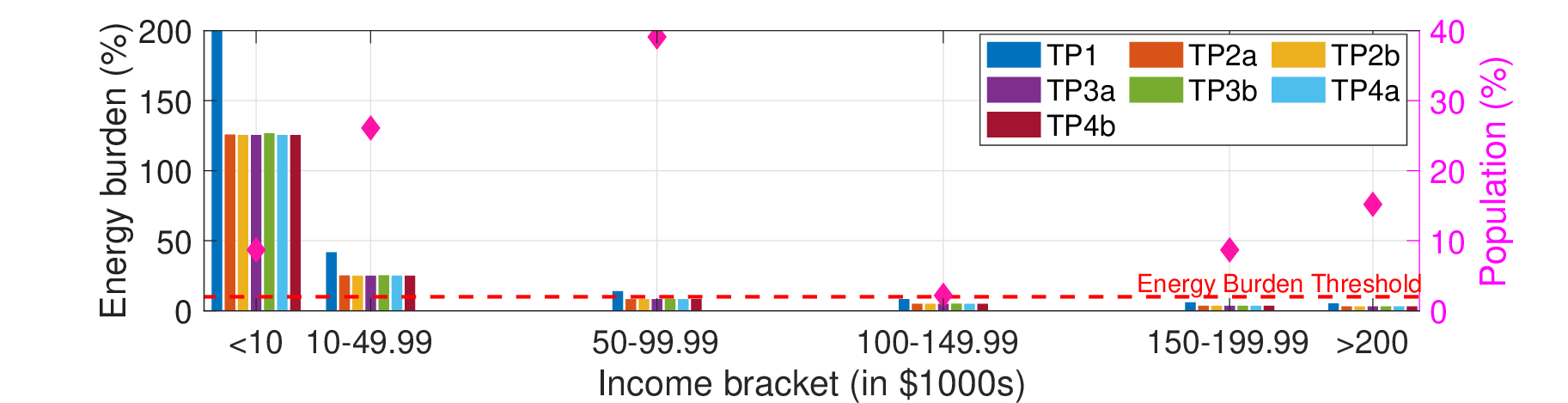}
    \caption{Variation in energy burden with respect to income bracket}
    \label{EB}
\end{figure}

Social metrics such as energy burden and energy poverty show considerable improvement with heat pump adoption. On average, households in Shungnak and Kobuk currently spend 19.8\% of their income on energy services. With heat pump adoption, this average energy burden decreases to 13.1\%. Figure~\ref{EB} shows the distribution of energy burden across income brackets for different transition scenarios. For each income group, energy burden is calculated as the ratio of the average household energy cost to the mean income of that group. In the base case scenario, over 70\% of the population experiences an energy burden, defined as energy payments exceeding 10\% of income. The highest burden occurs in households earning less than \$10k (8\% of the total population), with energy costs reaching up to 200\% of their income. Adoption of heat pumps significantly reduces the energy burden for this group. While heat pumps do not eliminate the energy burden of the entire community, their adoption decreases the proportion of the population experiencing it from 74\% to 40\%. Also, it reduces the severity of the burden among lower-income households, underscoring their potential to promote energy equity.

Energy poverty, estimated from the NANA survey based on the share of the population experiencing issues such as broken or malfunctioning furnaces, inability to afford fuel, or lack of access to firewood, also improves. In the base case (TP1), 40\% population experiences energy poverty, whereas this share decreases to 26\% under heat pump pathways. This reduction indicates an improvement in energy efficiency due to minimizing the dependency on a single heating source. With the deployment of heat pumps, furnaces are used only during the coldest periods (approximately 80 out of the 330 days when heating is needed).

Lastly, we calculated the transition scores to compare across the transition pathways, by equally weighting them across all criteria, using AHP, a widely adopted MCDA as described in Section~\ref{MCDA}. It reveals that the base case scenario (TP1) has the lowest transition score (0.42), while TP4b has the highest (0.75). However, it is important to note that infrastructure costs in TP4b do not account for the additional PV installation expenses, since it is covered by the Federal funds \citep{oced_news}. In general, 12 MBtu/h HP scenarios outperform their 18 MBtu/h counterparts. For instance, the TP2b (12 MBtu/h, uncoordinated) scenario functions 12\% 
Overall, pathways incorporating coordinated HPs achieve more balanced improvements across economic, environmental, and social metrics, whereas uncoordinated HP pathways primarily focus on targeted benefits, such as CO\textsubscript{2e} reductions. Importantly, varying the weights to focus more on specific dimensions of the TEES framework did not alter the relative rankings of the transition pathways, indicating the robustness of the evaluation outcomes.

{The successful transition towards renewable projects in remote communities is strongly supported by factors such as the availability of subsidies and robust partnerships among stakeholders, including utilities (e.g., AVEC), local governments (e.g., NAB), and the community itself \citep{holdmann2022critical}. Our findings are highly consistent with those presented by Chang and colleagues\citep{chang2022energy}, who evaluated the impact of rooftop solar PV and electric vehicles in Korea, demonstrating the broad applicability of these underlying principles for energy transitions.}

The household energy costs reported in Table~\ref{indices} for scenarios with heat pumps account for specific outdoor cutoff temperatures and indoor temperature settings (c.f table~\ref{HP_param}), pertaining to Arctic temperature settings. Since heat pump energy consumption is highly temperature-dependent, the following subsections present a sensitivity analysis examining how variations in indoor and outdoor cutoff temperatures affect cost savings, using TP2b as a case transition pathway.

\subsection{Impact of Heat Pump Cutoff Temperature on Cost Savings}
\begin{figure}[htbp!]
    \centering
    {\includegraphics[width=0.6\linewidth]{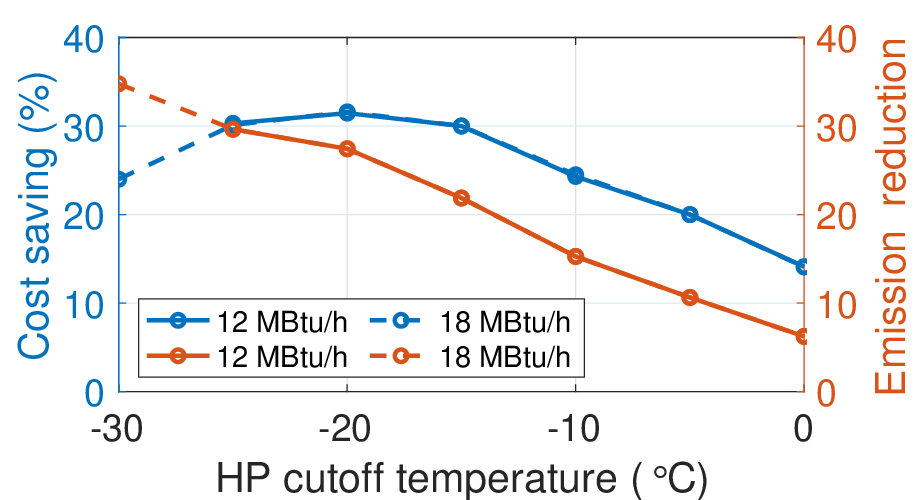}}
    \caption{{Variation in cost savings and CO\textsubscript{2e} emissions with respect to heat pump cutoff temperature}}
    \label{costsaving}
\end{figure}

Figure~\ref{costsaving} illustrates the relationship between heat pump cutoff temperature and CO\textsubscript{2e} emissions reductions and its related cost savings. The results indicate maximum cost savings when HPs are operated down to approximately $-20^\circ$C. Below this threshold, savings decline as additional infrastructure and higher electricity consumption offset the benefits of reduced fuel use. Conversely, CO\textsubscript{2e} reductions continue to improve with lower cutoff temperatures, peaking at $-25^\circ$C. This indicates a clear trade-off; lower cutoff temperatures enhance environmental benefits at the cost of economic efficiency.

This trade-off highlights the value of dual heating systems, where households retain both HPs and conventional heating systems, switching between them based on outdoor temperature to optimize cost and emissions. Uncoordinated and coordinated HP operation follow a similar trend, though coordinated operation consistently provides better economic outcomes due to load smoothing and reduced peak demand-related upgrades. 

However, the effectiveness of this dual-system approach is predicated on the operational status of existing conventional furnaces. As detailed in Table~\ref{HeatingSurvey}, many households in Shungnak-Kobuk report having broken or malfunctioning furnaces. For these households, the potential for maximum savings from a dual system is unrealized, contributing to their ongoing financial burden.
\begin{figure}[htbp!]
    \centering
    \includegraphics[width=0.6\linewidth]{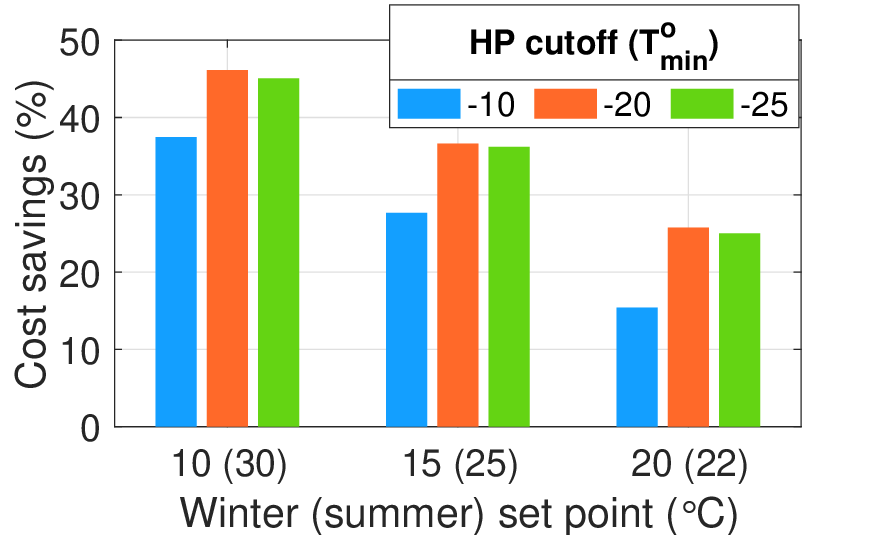}
    \caption{Cost savings under TP2b scenario as a function of indoor temperature set point}
    \label{fig18a}
\end{figure}
\subsection{Impact of Indoor Temperature Setpoint on Cost Savings}
{Figure~\ref{fig18a} illustrates the impact of heat pump indoor temperature setpoint adjustments on annual cost savings. It shows that lower temperature setpoints have higher savings.}
{Energy limiting behaviors may alleviate some immediate financial pressure, but it comes at the expense of thermal comfort, which reflects a form of energy poverty. In this context, households may lower setpoints out of financial necessity than choice \citep{cong2022unveiling}.}

{To address the energy equity gap, it is imperative to implement solutions that go beyond behavioral adjustments. The cost savings realized from HP adoption could be reallocated toward structural improvements that enhance long-term comfort and equity. This strategy creates a positive feedback loop; HP adoption not only lowers utility bills but also provides the capital to address the underlying structural deficiencies that are the root causes of energy poverty. HP deployment thus can increase energy efficiency, lower some socioeconomic pressures, and address equity concerns.}
 

{The analysis also revealed that short-term fluctuations in solar photovoltaic (PV) generation have a minimal impact on cost savings. For instance, a $\pm$20\% variation in PV availability results in only a $\pm$1\% change in household energy savings. This insensitivity is attributed to the fact that the electricity demand for a heat pump is primarily dictated by outdoor temperature and remains relatively stable, while the overall contribution of solar PV to the total system demand is comparatively small for Shungnak and Kobuk.}

\subsection{\tck{Generalizing Framework to Other Energy Communities}}
The optimization-based framework presented in this paper offers a generalizable methodology for assessing energy transition pathways in remote microgrids 
communities, extending beyond the specific case study of Shungnak and Kobuk. Its modular design, which integrates system modeling, multi-criteria decision analysis, and assessment, supports broad applicability across diverse microgrids.

The presented framework can be readily applied by collecting relevant data to different resource and electrification scenarios (e.g., wind, hydro, thermal storage, and electrified transportation), including network constraints. 
For remote communities in Alaska, most information is readily available from the PCE reports published by the Alaska Energy Authority. Other data, such as demand, renewable generation, transformer details, and network configurations, may not be easily accessible. This limitation does not affect the applicability of the framework itself; rather, it highlights the importance of a community collaborative approach.  Working with the utility, local government, and residents can help overcome this data gap through surveys, participatory mapping, and sharing operational knowledge, which is often not formally documented. Such an approach not only improves the data quality but also ensures that the modeling and decision analysis reflect community priorities. In cases where data is unavailable, reasonable assumptions can be made using standard sources such as utility reports, state and federal energy databases, and weather datasets, with sensitivity analyses performed to strengthen the results.



Since the paper's formulation is developed for remote communities and other off-grid microgrids, adaptation to more general grid-connected systems, such as rural or suburban distribution feeders, will require a non-trivial extension. For example, interactions between grid-connected communities and wholesale markets, IPPs, power purchase agreements (PPAs), change the concept of optimal energy dispatch since there is generally no single generator, but instead something akin to an \textit{slack bus} at the substation level. 
Incorporating these factors introduces challenges in economic, operational, and planning assumptions~\cite{NREL_HEM}. Addressing these challenges is beyond the scope of this study but represents an interesting direction for future work.


\section{Conclusion \& Future Work}
This study proposed a community-centric multi-criteria assessment framework to evaluate energy transition pathways. The framework provides an approach to assess infrastructure requirements, energy costs, emissions reductions, and social impacts for informed decision-making for sustainable energy transition.  

The framework was applied to the remote Alaskan communities of Shungnak and Kobuk. The three transition pathways considered for evaluation are heat pump and battery integration, intelligent coordination of resources, and expanded community solar PV capacity. The study indicated that heat pumps substantially reduce both CO\textsubscript{2} emissions and total energy costs. Coordinated heat pump operation further enhanced these benefits by minimizing peak loads and therefore reducing the need for costly infrastructure upgrades. Adoption of heat pumps also lowered annual household energy costs by up to 33\%. Additional solar PV integration improves these gains, achieving up to a 28\% 

The proposed assessment framework is designed to be flexible and scalable, allowing its application across different energy communities. Future work will focus on generalizing the framework to the grid-connected systems, accounting for variations in renewable generation, electricity price, demand, etc. This will involve incorporating uncertainties through scenario-based optimization and sensitivity analysis to further strengthen the methodology. 

\section*{Acknowledgment}
This work was supported by the generous funds from the Sloan Foundation and the Catalyst Award from the Gund Institute of the Environment at the University of Vermont. We are profoundly grateful to the Tribe of Shungnak and the Tribe of Kobuk, where this research is based. This work would not have been possible without the support of Billy Lee from Shungnak, Darren Westby, and Bill Stamm at AVEC, the NANA Regional Corporation, and the Northwest Arctic Borough, who provided the data to do this analysis, and for their generous guidance and ongoing feedback on this work. Additionally, we are grateful for our monthly debriefings with our collaborators, Michelle Wilbur, Shivani Bhagat, and Christie Haupert at the Alaska Center for Energy and Power at the University of Alaska Fairbanks.

 \bibliographystyle{elsarticle-num-names} 
 \bibliography{References}

\begin{thebibliography}{47}
\expandafter\ifx\csname natexlab\endcsname\relax\def\natexlab#1{#1}\fi
\providecommand{\url}[1]{\texttt{#1}}
\providecommand{\href}[2]{#2}
\providecommand{\path}[1]{#1}
\providecommand{\DOIprefix}{doi:}
\providecommand{\ArXivprefix}{arXiv:}
\providecommand{\URLprefix}{URL: }
\providecommand{\Pubmedprefix}{pmid:}
\providecommand{\doi}[1]{\href{http://dx.doi.org/#1}{\path{#1}}}
\providecommand{\Pubmed}[1]{\href{pmid:#1}{\path{#1}}}
\providecommand{\bibinfo}[2]{#2}
\ifx\xfnm\relax \def\xfnm[#1]{\unskip,\space#1}\fi
\bibitem[{Gielen et~al.(2019)Gielen, Boshell, Saygin, Bazilian, Wagner, and Gorini}]{gielen2019role}
\bibinfo{author}{D.~Gielen}, \bibinfo{author}{F.~Boshell}, \bibinfo{author}{D.~Saygin}, \bibinfo{author}{M.~D. Bazilian}, \bibinfo{author}{N.~Wagner}, \bibinfo{author}{R.~Gorini},
\newblock \bibinfo{title}{The role of renewable energy in the global energy transformation},
\newblock \bibinfo{journal}{Energy strategy reviews} \bibinfo{volume}{24} (\bibinfo{year}{2019}) \bibinfo{pages}{38--50}.
\bibitem[{McKenna et~al.(2018)McKenna, Bchini, Weinand, Michaelis, K{\"o}nig, K{\"o}ppel, and Fichtner}]{mckenna2018future}
\bibinfo{author}{R.~C. McKenna}, \bibinfo{author}{Q.~Bchini}, \bibinfo{author}{J.~M. Weinand}, \bibinfo{author}{J.~Michaelis}, \bibinfo{author}{S.~K{\"o}nig}, \bibinfo{author}{W.~K{\"o}ppel}, \bibinfo{author}{W.~Fichtner},
\newblock \bibinfo{title}{The future role of power-to-gas in the energy transition: Regional and local techno-economic analyses in baden-w{\"u}rttemberg},
\newblock \bibinfo{journal}{Applied energy} \bibinfo{volume}{212} (\bibinfo{year}{2018}) \bibinfo{pages}{386--400}.
\bibitem[{Ibagon et~al.(2023)Ibagon, Mu{\~n}oz, and Correa}]{ibagon2023techno}
\bibinfo{author}{N.~Ibagon}, \bibinfo{author}{P.~Mu{\~n}oz}, \bibinfo{author}{G.~Correa},
\newblock \bibinfo{title}{Techno economic analysis tool for the sizing and optimization of an off-grid hydrogen hub},
\newblock \bibinfo{journal}{Journal of Energy Storage} \bibinfo{volume}{73} (\bibinfo{year}{2023}) \bibinfo{pages}{108787}.
\bibitem[{Bhandari and Shah(2021)}]{bhandari2021hydrogen}
\bibinfo{author}{R.~Bhandari}, \bibinfo{author}{R.~R. Shah},
\newblock \bibinfo{title}{Hydrogen as energy carrier: Techno-economic assessment of decentralized hydrogen production in germany},
\newblock \bibinfo{journal}{Renewable Energy} \bibinfo{volume}{177} (\bibinfo{year}{2021}) \bibinfo{pages}{915--931}.
\bibitem[{Osei et~al.(2024)Osei, Odoi-Yorke, Opoku, Baah, Obeng, Mensah, and Forson}]{osei2024techno}
\bibinfo{author}{L.~K. Osei}, \bibinfo{author}{F.~Odoi-Yorke}, \bibinfo{author}{R.~Opoku}, \bibinfo{author}{B.~Baah}, \bibinfo{author}{G.~Y. Obeng}, \bibinfo{author}{L.~D. Mensah}, \bibinfo{author}{F.~K. Forson},
\newblock \bibinfo{title}{Techno-economic viability of decentralised solar photovoltaic-based green hydrogen production for sustainable energy transition in ghana},
\newblock \bibinfo{journal}{Solar Compass} \bibinfo{volume}{9} (\bibinfo{year}{2024}) \bibinfo{pages}{100068}.
\bibitem[{Chang et~al.(2022)Chang, Cho, Heo, Kang, and Kobashi}]{chang2022energy}
\bibinfo{author}{S.~Chang}, \bibinfo{author}{J.~Cho}, \bibinfo{author}{J.~Heo}, \bibinfo{author}{J.~Kang}, \bibinfo{author}{T.~Kobashi},
\newblock \bibinfo{title}{Energy infrastructure transitions with pv and ev combined systems using techno-economic analyses for decarbonization in cities},
\newblock \bibinfo{journal}{Applied Energy} \bibinfo{volume}{319} (\bibinfo{year}{2022}) \bibinfo{pages}{119254}.
\bibitem[{Van~Dael et~al.(2013)Van~Dael, Van~Passel, Pelkmans, Guisson, Reumermann, Luzardo, Witters, and Broeze}]{van2013techno}
\bibinfo{author}{M.~Van~Dael}, \bibinfo{author}{S.~Van~Passel}, \bibinfo{author}{L.~Pelkmans}, \bibinfo{author}{R.~Guisson}, \bibinfo{author}{P.~Reumermann}, \bibinfo{author}{N.~M. Luzardo}, \bibinfo{author}{N.~Witters}, \bibinfo{author}{J.~Broeze},
\newblock \bibinfo{title}{A techno-economic evaluation of a biomass energy conversion park},
\newblock \bibinfo{journal}{Applied energy} \bibinfo{volume}{104} (\bibinfo{year}{2013}) \bibinfo{pages}{611--622}.
\bibitem[{Cherp et~al.(2018)Cherp, Vinichenko, Jewell, Brutschin, and Sovacool}]{cherp2018integrating}
\bibinfo{author}{A.~Cherp}, \bibinfo{author}{V.~Vinichenko}, \bibinfo{author}{J.~Jewell}, \bibinfo{author}{E.~Brutschin}, \bibinfo{author}{B.~Sovacool},
\newblock \bibinfo{title}{Integrating techno-economic, socio-technical and political perspectives on national energy transitions: A meta-theoretical framework},
\newblock \bibinfo{journal}{Energy Research \& Social Science} \bibinfo{volume}{37} (\bibinfo{year}{2018}) \bibinfo{pages}{175--190}.
\bibitem[{Bolwig et~al.(2020)Bolwig, Bolkesj{\o}, Klitkou, Lund, Bergaentzl{\'e}, Borch, Olsen, Kirkerud, Chen, Gunkel et~al.}]{bolwig2020climate}
\bibinfo{author}{S.~Bolwig}, \bibinfo{author}{T.~F. Bolkesj{\o}}, \bibinfo{author}{A.~Klitkou}, \bibinfo{author}{P.~D. Lund}, \bibinfo{author}{C.~Bergaentzl{\'e}}, \bibinfo{author}{K.~Borch}, \bibinfo{author}{O.~J. Olsen}, \bibinfo{author}{J.~G. Kirkerud}, \bibinfo{author}{Y.-k. Chen}, \bibinfo{author}{P.~A. Gunkel}, et~al.,
\newblock \bibinfo{title}{Climate-friendly but socially rejected energy-transition pathways: The integration of techno-economic and socio-technical approaches in the nordic-baltic region},
\newblock \bibinfo{journal}{Energy Research \& Social Science} \bibinfo{volume}{67} (\bibinfo{year}{2020}) \bibinfo{pages}{101559}.
\bibitem[{Heras and Mart{\'\i}n(2020)}]{heras2020social}
\bibinfo{author}{J.~Heras}, \bibinfo{author}{M.~Mart{\'\i}n},
\newblock \bibinfo{title}{Social issues in the energy transition: Effect on the design of the new power system},
\newblock \bibinfo{journal}{Applied Energy} \bibinfo{volume}{278} (\bibinfo{year}{2020}) \bibinfo{pages}{115654}.
\bibitem[{Garcia-Casals et~al.(2019)Garcia-Casals, Ferroukhi, and Parajuli}]{garcia2019measuring}
\bibinfo{author}{X.~Garcia-Casals}, \bibinfo{author}{R.~Ferroukhi}, \bibinfo{author}{B.~Parajuli},
\newblock \bibinfo{title}{Measuring the socio-economic footprint of the energy transition},
\newblock \bibinfo{journal}{Energy Transitions} \bibinfo{volume}{3} (\bibinfo{year}{2019}) \bibinfo{pages}{105--118}.
\bibitem[{Tladi et~al.(2024)Tladi, Kambule, and Modley}]{tladi2024assessing}
\bibinfo{author}{B.~Tladi}, \bibinfo{author}{N.~Kambule}, \bibinfo{author}{L.-A. Modley},
\newblock \bibinfo{title}{Assessing the social and environmental impacts of the just energy transition in komati, mpumalanga province, south africa},
\newblock \bibinfo{journal}{Energy Research \& Social Science} \bibinfo{volume}{111} (\bibinfo{year}{2024}) \bibinfo{pages}{103489}.
\bibitem[{Vanegas-Cantarero et~al.(2022)Vanegas-Cantarero, Pennock, Bloise-Thomaz, Jeffrey, and Dickson}]{vanegas2022beyond}
\bibinfo{author}{M.~M. Vanegas-Cantarero}, \bibinfo{author}{S.~Pennock}, \bibinfo{author}{T.~Bloise-Thomaz}, \bibinfo{author}{H.~Jeffrey}, \bibinfo{author}{M.~J. Dickson},
\newblock \bibinfo{title}{Beyond lcoe: A multi-criteria evaluation framework for offshore renewable energy projects},
\newblock \bibinfo{journal}{Renewable and Sustainable Energy Reviews} \bibinfo{volume}{161} (\bibinfo{year}{2022}) \bibinfo{pages}{112307}.
\bibitem[{Zhang and Kong(2022)}]{zhang2022green}
\bibinfo{author}{D.~Zhang}, \bibinfo{author}{Q.~Kong},
\newblock \bibinfo{title}{Green energy transition and sustainable development of energy firms: An assessment of renewable energy policy},
\newblock \bibinfo{journal}{Energy Economics} \bibinfo{volume}{111} (\bibinfo{year}{2022}) \bibinfo{pages}{106060}.
\bibitem[{Markard(2018)}]{markard2018next}
\bibinfo{author}{J.~Markard},
\newblock \bibinfo{title}{The next phase of the energy transition and its implications for research and policy},
\newblock \bibinfo{journal}{Nature Energy} \bibinfo{volume}{3} (\bibinfo{year}{2018}) \bibinfo{pages}{628--633}.
\bibitem[{Li et~al.(2025)Li, Niu, Wang, Tan, and Wang}]{li2025multidimensional}
\bibinfo{author}{Z.~Li}, \bibinfo{author}{S.~Niu}, \bibinfo{author}{J.~Wang}, \bibinfo{author}{Y.~Tan}, \bibinfo{author}{Z.~Wang},
\newblock \bibinfo{title}{Multidimensional assessment of energy transition and policy implications},
\newblock \bibinfo{journal}{Renewable Energy} \bibinfo{volume}{238} (\bibinfo{year}{2025}) \bibinfo{pages}{121870}.
\bibitem[{Holdmann et~al.(2022)Holdmann, Pride, Poelzer, Noble, and Walker}]{holdmann2022critical}
\bibinfo{author}{G.~Holdmann}, \bibinfo{author}{D.~Pride}, \bibinfo{author}{G.~Poelzer}, \bibinfo{author}{B.~Noble}, \bibinfo{author}{C.~Walker},
\newblock \bibinfo{title}{Critical pathways to renewable energy transitions in remote alaska communities: A comparative analysis},
\newblock \bibinfo{journal}{Energy Research \& Social Science} \bibinfo{volume}{91} (\bibinfo{year}{2022}) \bibinfo{pages}{102712}. \DOIprefix\doi{https://doi.org/10.1016/j.erss.2022.102712}.
\bibitem[{Allen et~al.(2016)Allen, Brutkoski, Farnsworth, and Larsen}]{allen2016sustainable}
\bibinfo{author}{R.~Allen}, \bibinfo{author}{D.~Brutkoski}, \bibinfo{author}{D.~Farnsworth}, \bibinfo{author}{P.~Larsen},
\newblock \bibinfo{title}{Sustainable energy solutions for rural alaska}  (\bibinfo{year}{2016}).
\bibitem[{Heleno et~al.(2024)Heleno, Brown, Valenzuela, Orrell, Sheridan, Kazimierczuk, Hudson, Gunda, and Beshilas}]{heleno2024resilient}
\bibinfo{author}{M.~Heleno}, \bibinfo{author}{R.~Brown}, \bibinfo{author}{A.~Valenzuela}, \bibinfo{author}{A.~Orrell}, \bibinfo{author}{L.~Sheridan}, \bibinfo{author}{K.~Kazimierczuk}, \bibinfo{author}{H.~Hudson}, \bibinfo{author}{T.~Gunda}, \bibinfo{author}{L.~Beshilas}, \bibinfo{title}{Resilient Energy Transition Planning for Ouzinkie, Alaska}, \bibinfo{type}{Technical Report}, National Renewable Energy Laboratory (NREL), Golden, CO (United States), \bibinfo{year}{2024}.
\bibitem[{Trueworthy et~al.(2024)Trueworthy, McCarrel, Wieliczkiewicz, Cellan, Peterson, Anderson, DuPont, and Grear}]{trueworthy2024will}
\bibinfo{author}{A.~Trueworthy}, \bibinfo{author}{A.~McCarrel}, \bibinfo{author}{J.~Wieliczkiewicz}, \bibinfo{author}{S.~Cellan}, \bibinfo{author}{W.~Peterson}, \bibinfo{author}{S.~Anderson}, \bibinfo{author}{B.~DuPont}, \bibinfo{author}{M.~Grear},
\newblock \bibinfo{title}{Who will be making wave energy? a community-driven design approach toward just and sustainable energy futures in alaska},
\newblock \bibinfo{journal}{Energy Research \& Social Science} \bibinfo{volume}{115} (\bibinfo{year}{2024}) \bibinfo{pages}{103615}.
\bibitem[{Arriaga et~al.(2013)Arriaga, Ca{\~n}izares, and Kazerani}]{arriaga2013renewable}
\bibinfo{author}{M.~Arriaga}, \bibinfo{author}{C.~A. Ca{\~n}izares}, \bibinfo{author}{M.~Kazerani},
\newblock \bibinfo{title}{Renewable energy alternatives for remote communities in northern ontario, canada},
\newblock \bibinfo{journal}{IEEE Transactions on sustainable energy} \bibinfo{volume}{4} (\bibinfo{year}{2013}) \bibinfo{pages}{661--670}.
\bibitem[{{Northwest Arctic Borough Alaska}(2022)}]{nwab2022}
\bibinfo{author}{{Northwest Arctic Borough Alaska}}, \bibinfo{title}{Northwest arctic regional energy plan 2022}, \bibinfo{type}{Technical Report}, \bibinfo{year}{2022}. \URLprefix \url{https://www.nwabor.org/wp-content/uploads/NWAB-Regional-Energy-Plan-Update-Final.pdf}.
\bibitem[{Wang et~al.(2009)Wang, Jing, Zhang, and Zhao}]{wang2009review}
\bibinfo{author}{J.-J. Wang}, \bibinfo{author}{Y.-Y. Jing}, \bibinfo{author}{C.-F. Zhang}, \bibinfo{author}{J.-H. Zhao},
\newblock \bibinfo{title}{Review on multi-criteria decision analysis aid in sustainable energy decision-making},
\newblock \bibinfo{journal}{Renewable and sustainable energy reviews} \bibinfo{volume}{13} (\bibinfo{year}{2009}) \bibinfo{pages}{2263--2278}.
\bibitem[{Kumar et~al.(2017)Kumar, Sah, Singh, Deng, He, Kumar, and Bansal}]{kumar2017review}
\bibinfo{author}{A.~Kumar}, \bibinfo{author}{B.~Sah}, \bibinfo{author}{A.~R. Singh}, \bibinfo{author}{Y.~Deng}, \bibinfo{author}{X.~He}, \bibinfo{author}{P.~Kumar}, \bibinfo{author}{R.~C. Bansal},
\newblock \bibinfo{title}{A review of multi criteria decision making (mcdm) towards sustainable renewable energy development},
\newblock \bibinfo{journal}{Renewable and sustainable energy reviews} \bibinfo{volume}{69} (\bibinfo{year}{2017}) \bibinfo{pages}{596--609}.
\bibitem[{Rossol et~al.(2018)Rossol, Michael, Brinkman, Buster, Denholm, Novacheck, and Stephen}]{nrel_data}
\bibinfo{author}{Rossol}, \bibinfo{author}{G.~Michael}, \bibinfo{author}{G.~Brinkman}, \bibinfo{author}{P.~Buster}, \bibinfo{author}{J.~Denholm}, \bibinfo{author}{G.~Novacheck}, \bibinfo{author}{Stephen}, \bibinfo{title}{A National Thermal Generator Performance Database}, \bibinfo{publisher}{NREL Data Catalog. Golden, CO: National Renewable Energy Laboratory}, \bibinfo{year}{2018}. \URLprefix \url{https://data.nrel.gov/submissions/100}. \DOIprefix\doi{10.7799/1499030}.
\bibitem[{Bergen and Vittal(2000)}]{bergen_power_2000}
\bibinfo{author}{A.~Bergen}, \bibinfo{author}{V.~Vittal}, \bibinfo{title}{Power {Systems} {Analysis}}, \bibinfo{publisher}{Prentice Hall}, \bibinfo{year}{2000}. \URLprefix \url{https://books.google.com/books?id=4InAQwAACAAJ}.
\bibitem[{Vykhodtsev et~al.(2022)Vykhodtsev, Jang, Wang, Rosehart, and Zareipour}]{vykhodtsev2022review}
\bibinfo{author}{A.~V. Vykhodtsev}, \bibinfo{author}{D.~Jang}, \bibinfo{author}{Q.~Wang}, \bibinfo{author}{W.~Rosehart}, \bibinfo{author}{H.~Zareipour},
\newblock \bibinfo{title}{A review of modelling approaches to characterize lithium-ion battery energy storage systems in techno-economic analyses of power systems},
\newblock \bibinfo{journal}{Renewable and Sustainable Energy Reviews} \bibinfo{volume}{166} (\bibinfo{year}{2022}) \bibinfo{pages}{112584}.
\bibitem[{Oyefeso et~al.(2022)Oyefeso, Ledva, Almassalkhi, Hiskens, and Mathieu}]{oyefeso_control_2022}
\bibinfo{author}{O.~Oyefeso}, \bibinfo{author}{G.~S. Ledva}, \bibinfo{author}{M.~Almassalkhi}, \bibinfo{author}{I.~A. Hiskens}, \bibinfo{author}{J.~L. Mathieu},
\newblock \bibinfo{title}{Control of {Aggregate} {Air}-{Conditioning} {Load} using {Packetized} {Energy} {Concepts}},
\newblock in: \bibinfo{booktitle}{2022 {IEEE} {Conference} on {Control} {Technology} and {Applications} ({CCTA})}, \bibinfo{year}{2022}, pp. \bibinfo{pages}{447--454}. \DOIprefix\doi{10.1109/CCTA49430.2022.9966155}, \bibinfo{note}{iSSN: 2768-0770}.
\bibitem[{ana(2018)}]{analysisnorth}
\bibinfo{title}{Mini-split heat pumps in alaska: Heat pump calculator algorithms and data}, \bibinfo{year}{2018}. \URLprefix \url{https://docs.google.com/document/u/0/d/1jLZ2JBw1Zj40W7y7QrZPMUknNrksk6y4NZlHD_2Fs9E/preview?hgd=1&usp=embed_facebook}.
\bibitem[{Baker(2019)}]{baker2019}
\bibinfo{author}{S.~H. Baker},
\newblock \bibinfo{title}{Anti-resilience: a roadmap for transformational justice within the energy system},
\newblock \bibinfo{journal}{Harv. CR-CLL Rev.} \bibinfo{volume}{54} (\bibinfo{year}{2019}) \bibinfo{pages}{1}.
\bibitem[{Ma et~al.(2019)Ma, Laymon, Day, Oliveira, Weers, and Vimont}]{EB_def}
\bibinfo{author}{O.~Ma}, \bibinfo{author}{K.~Laymon}, \bibinfo{author}{M.~Day}, \bibinfo{author}{R.~Oliveira}, \bibinfo{author}{J.~Weers}, \bibinfo{author}{A.~Vimont}, \bibinfo{title}{Low-{Income} {Energy} {Affordability} {Data} ({LEAD}) {Tool} {Methodology}}, \bibinfo{type}{Technical Report} \bibinfo{number}{NREL/TP--6A20-74249, 1545589}, \bibinfo{year}{2019}. \URLprefix \url{https://www.osti.gov/servlets/purl/1545589/}. \DOIprefix\doi{10.2172/1545589}.
\bibitem[{Vafaei et~al.(2016)Vafaei, Ribeiro, and Camarinha-Matos}]{vafaei2016normalization}
\bibinfo{author}{N.~Vafaei}, \bibinfo{author}{R.~A. Ribeiro}, \bibinfo{author}{L.~M. Camarinha-Matos},
\newblock \bibinfo{title}{Normalization techniques for multi-criteria decision making: analytical hierarchy process case study},
\newblock in: \bibinfo{booktitle}{Technological innovation for cyber-physical systems: 7th IFIP WG 5.5/SOCOLNET advanced doctoral conference on computing, electrical and industrial systems, DoCEIS 2016, Costa de Caparica, Portugal, April 11--13, 2016, Proceedings 7}, \bibinfo{organization}{Springer}, \bibinfo{year}{2016}, pp. \bibinfo{pages}{261--269}.
\bibitem[{{Northwest Arctic Borough Alaska}(2022{\natexlab{a}})}]{shungnak}
\bibinfo{author}{{Northwest Arctic Borough Alaska}}, \bibinfo{title}{Community profile shungnak}, \bibinfo{year}{2022}{\natexlab{a}}. \URLprefix \url{https://www.nwabor.org/wp-content/uploads/L.-Community-Profile-Shungnak.pdf}.
\bibitem[{{Northwest Arctic Borough Alaska}(2022{\natexlab{b}})}]{kobuk}
\bibinfo{author}{{Northwest Arctic Borough Alaska}}, \bibinfo{title}{Community profile kobuk}, \bibinfo{year}{2022}{\natexlab{b}}. \URLprefix \url{https://www.nwabor.org/wp-content/uploads/G.-Community-Profile-Kobuk.pdf}.
\bibitem[{Devine and Baring-Gould(2004)}]{devine2004alaska}
\bibinfo{author}{M.~Devine}, \bibinfo{author}{E.~I. Baring-Gould}, \bibinfo{title}{Alaska Village Electric Load Calculator}, \bibinfo{type}{Technical Report}, Citeseer, \bibinfo{year}{2004}.
\bibitem[{{Ageto Energy}(2021)}]{shungnak_2021}
\bibinfo{author}{{Ageto Energy}}, \bibinfo{year}{2021}. \URLprefix \url{https://agetoenergy.com/case_studies/shungnak-alaska/}.
\bibitem[{{Alaska Energy Authority}(2024)}]{pce}
\bibinfo{author}{{Alaska Energy Authority}}, \bibinfo{title}{{FY23} {PCE} statistical report}, \bibinfo{year}{2024}. \URLprefix \url{https://www.akenergyauthority.org/Portals/0/Power%20Cost%20Equalization/2024.02.26%20FY23%20PCE%20Statistical%20Report%20by%20Community%20(Final%20Optimzed).pdf?ver=om4p4ZK_A-xwHiFPOHfvDQ%3d%3d}.
\bibitem[{{United States Department of Transportation, Bureau of Transportation Statistics}(2019)}]{diesel_density}
\bibinfo{author}{{United States Department of Transportation, Bureau of Transportation Statistics}}, \bibinfo{title}{National {Transportation} {Statistics} ({NTS})}, \bibinfo{year}{2019}. \URLprefix \url{https://tinyurl.com/rosapntlbtsNTS}. \DOIprefix\doi{10.21949/1503663}.
\bibitem[{DeMarban(2024)}]{ADN2024HeatPumps}
\bibinfo{author}{A.~DeMarban},
\newblock \bibinfo{title}{In northwest arctic, federal grant will bring heat pumps for households, solar energy in villages},
\newblock \bibinfo{journal}{Anchorage Daily News}  (\bibinfo{year}{2024}). \URLprefix \url{https://www.adn.com/business-economy/energy/2024/03/05/in-northwest-arctic-federal-grant-will-bring-heat-pumps-for-households-solar-energy-in-villages/}.
\bibitem[{{Northwest Arctic Borough Alaska}(2022)}]{ambler}
\bibinfo{author}{{Northwest Arctic Borough Alaska}}, \bibinfo{title}{Community profile ambler}, \bibinfo{year}{2022}. \URLprefix \url{https://www.nwabor.org/wp-content/uploads/B.-Community-Profile-Ambler.pdf}.
\bibitem[{Jeremy(2024)}]{jeremy_how_2024}
\bibinfo{author}{Jeremy}, \bibinfo{title}{How {Much} {Does} a {Power} {Pole} {Transformer} {Cost}}, \bibinfo{year}{2024}. \URLprefix \url{https://evernewtransformer.com/how-much-does-a-power-pole-transformer-cost/}.
\bibitem[{lar(2025)}]{larson_nodate}
\bibinfo{title}{Larson {Electronics} - 250 {KVA} {Pad} {Mount} {Transformer} - {12470Y}/7200 {Wye}-{N} {Primary}, 120/{240V} {Delta} {Secondary} - {Copper}, {ONAN}/{Bell} {Green}}, \bibinfo{year}{2025}. \URLprefix \url{{https://www.larsonelectronics.com/product/305111/250-kva-pad-mount-transformer-12470y-7200-wye-n-primary-120-240v-delta-secondary-copper-onan-bell-green?productIdBase64=MzA1MTEx0&rank=18}}.
\bibitem[{AEA(2024)}]{AEA_pce_webpage}
\bibinfo{title}{Alaska {Energy} {Authority} {Power} {Cost} {Equalization}}, \bibinfo{year}{2024}. \URLprefix \url{https://www.akenergyauthority.org/What-We-Do/Power-Cost-Equalization}.
\bibitem[{{US EPA OAR}(2020)}]{ap42}
\bibinfo{author}{{US EPA OAR}}, \bibinfo{title}{{AP} 42, {Fifth} {Edition}, {Volume} {I} {Chapter} 3: {Stationary} {Internal} {Combustion} {Sources}}, \bibinfo{year}{2020}. \URLprefix \url{https://www.epa.gov/air-emissions-factors-and-quantification/ap-42-fifth-edition-volume-i-chapter-3-stationary-0}.
\bibitem[{{U.S. Department of Energy, Office of Clean Energy Demonstrations}(2024)}]{oced_news}
\bibinfo{author}{{U.S. Department of Energy, Office of Clean Energy Demonstrations}}, \bibinfo{title}{{Award Wednesdays, November 20, 2024}}, \bibinfo{howpublished}{\url{https://www.energy.gov/oced/articles/award-wednesdays-november-20-2024}}, \bibinfo{year}{2024}. \bibinfo{note}{[Accessed July 23, 2025]}.
\bibitem[{Cong et~al.(2022)Cong, Nock, Qiu, and Xing}]{cong2022unveiling}
\bibinfo{author}{S.~Cong}, \bibinfo{author}{D.~Nock}, \bibinfo{author}{Y.~L. Qiu}, \bibinfo{author}{B.~Xing},
\newblock \bibinfo{title}{Unveiling hidden energy poverty using the energy equity gap},
\newblock \bibinfo{journal}{Nature communications} \bibinfo{volume}{13} (\bibinfo{year}{2022}) \bibinfo{pages}{2456}.
\bibitem[{Guo et~al.(2023)Guo, Hale, Lau, Heeter, O'Malley, Palmintier, and Pless}]{NREL_HEM}
\bibinfo{author}{N.~Guo}, \bibinfo{author}{E.~Hale}, \bibinfo{author}{J.~Lau}, \bibinfo{author}{J.~Heeter}, \bibinfo{author}{M.~O'Malley}, \bibinfo{author}{B.~Palmintier}, \bibinfo{author}{S.~Pless},
\newblock \bibinfo{title}{Integrated, multi-stakeholder analysis of electricity system structures: Methodology and a case study},
\newblock \bibinfo{journal}{IEEE Transactions on Energy Markets, Policy and Regulation} \bibinfo{volume}{1} (\bibinfo{year}{2023}) \bibinfo{pages}{237--247}.

\end{thebibliography}

\end{document}